\RequirePackage{fix-cm}
\documentclass[fleqn,usenatbib]{mnras}
\usepackage{newtxtext,newtxmath}
\usepackage[T1]{fontenc}
\usepackage{graphicx,xcolor}
\usepackage{xspace}
\usepackage{booktabs,siunitx,multirow}

\newcommand{\posx}{\ensuremath{\boldsymbol{x}}\xspace}
\newcommand{\lam}{\ensuremath{\lambda}\xspace}
\newcommand{\flam}{\ensuremath{F_\lambda}\xspace}

\title[Continuous spatio-spectral sky]{Spectro-Spatial Survey Fusion}
\author[G. N. Vassilakis et al.]{Georgios N. Vassilakis$^{1,2,3}$,
Miles Cranmer$^{1,2,3}$, and
Vasily Belokurov$^{1,3}$\\
$^{1}$Institute of Astronomy, University of Cambridge, Cambridge, United Kingdom\\
$^{2}$Department of Applied Mathematics and Theoretical Physics, University of Cambridge, Cambridge, United Kingdom\\
$^{3}$Kavli Institute for Cosmology, University of Cambridge, Cambridge, United Kingdom
}
\date{}
\pubyear{\the\year}

\begin{document}
\label{firstpage}
\pagerange{\pageref{firstpage}--\pageref{lastpage}}
\maketitle

\begin{abstract}
    Imaging and spectroscopy constrain complementary information about the same sky but are usually processed separately. We introduce Spectro-Spatial Survey Fusion, a framework that unifies these measurements by fitting a continuous spectral surface-brightness field jointly through each instrument's response. Adapting from Gaussian splatting in computer vision, we represent the field as a sum of adaptive spatial Gaussians, each carrying a spectrum built from a shared, learned dictionary of templates. We demonstrate this on 400 Sloan Digital Sky Survey (SDSS) $ugriz$ exposures and 800 fiber spectra across an $84^{\prime}\times84^{\prime}$ region of COSMOS. The joint fit closely reproduces the spatial structure and broad spectral diversity of galaxies, stars, and quasars. For fitted SDSS observations, the summed predicted flux in source-selected pixels agrees with the data to within 2~percent in every band, and 96~percent of fitted spectra agree to within 10~percent in integrated flux. Sparse fiber spectra constrain a dictionary shared by every Gaussian component, so a model spectrum can be queried anywhere in the imaged area, including where no spectrum was fitted. Without refitting, the field predicts withheld SDSS images and spectra, and selected Hyper Suprime-Cam (HSC) images and Dark Energy Spectroscopic Instrument (DESI) spectra, which never enter the fit. The field predicts broad spiral-arm patterns in HSC images of two selected galaxies and DESI continuum shapes for galaxies without a fitted fiber. This framework provides a route to combining surveys so that spatial and spectral information from different observations and instruments can constrain a common representation of the sky.
\end{abstract}


\section{Introduction}
\label{sec:introduction}

Astronomical surveys provide complementary views of the same sky. Wide-field imaging programs, including the Dark Energy Spectroscopic Instrument (DESI) Legacy Imaging Surveys, Hyper Suprime-Cam (HSC), Rubin, Euclid, and Roman, differ in depth, angular resolution, and wavelength coverage \citep{dey2019legacy,aihara2022hscpdr3,ivezic2019lsst,euclid2025overview,spergel2015wfirst}. Multiplexed fiber spectrographs such as DESI, 4MOST, WEAVE, and PFS add detailed spectra for selected sources \citep{desi2016science,dejong2019fourmost,jin2024weave,takada2014pfs}. Combining these data requires reconciling their instrument responses as well as their source catalogs. Coaddition combines exposures on a common pixel grid within a band \citep{fruchter2002drizzle,zackay2017coadd1,zackay2017coadd2}, but the result retains a particular bandpass and effective point-spread function (PSF), and cannot directly incorporate a fiber spectrum. Matching catalogs has a different limitation: neighboring objects often overlap in deep images, so catalogs constructed independently at different resolutions need not agree on which sources are present \citep{bosch2018hscpipeline,melchior2018scarlet}. Joint processing at the pixel level allows information from one survey to help constrain measurements in another \citep{chary2020jsp}. Fiber spectroscopy samples only a fraction of the imaged sources, and each fiber collects a seeing-dependent portion of its target's light, potentially mixed with emission from neighbors. Fiber spectra are therefore already tied to imaging through models: target selection uses fiber fluxes predicted from imaging \citep{hahn2023bgs}, and aperture corrections use imaging to extend star-formation rates measured within a fiber to the whole galaxy \citep{brinchmann2004}. Slitless spectroscopy, as in the Euclid and Roman survey designs, records dispersed images whose overlaps must be modeled in both position and wavelength. Combining separately derived survey products does not, by itself, ensure that a single sky model can account for all the underlying measurements.

We instead model the measurements themselves. Under a static-sky approximation, calibrated image pixels and spectral bins can both be modeled as projections of the same underlying quantity: the spectral surface brightness of the sky $F_\lambda(\boldsymbol{x},\lambda)$, a function of sky position $\boldsymbol{x}$ and wavelength $\lambda$. An image pixel integrates $F_\lambda$ against a bandpass throughput, a PSF, and a pixel footprint; a fiber spectral bin integrates the same field against a narrow wavelength window, a line-spread function (LSF), and the fiber aperture. Rather than combining data products after the fact, we fit a continuous model of $F_\lambda$ directly to image pixels and spectral bins simultaneously, pushing the model through each instrument's own forward operator. Each additional instrument then contributes further constraints on the same latent field, which can be queried through different bandpasses, PSFs, apertures, and wavelength grids within the represented domain.

Fitting a common model through different instrument responses is already established in astronomical forward modeling. The Tractor and The Farmer connect photometric measurements across surveys through source models \citep{lang2016tractor,weaver2023farmer}, while scarlet combines morphology and color across bands to separate overlapping sources \citep{melchior2018scarlet}. Joint image reconstruction can account for different PSFs and pixel scales, allowing observations at different resolutions to constrain the same scene \citep{joseph2021jsp,donath2024jolideco}. Hyperspectral fusion reconstructs a latent spatio-spectral cube from imaging and integral-field unit (IFU) spectroscopy covering the same region \citep{guilloteau2020fusion}. \citet{marquis2026fusion} demonstrate this with James Webb Space Telescope (JWST) observations of a protoplanetary disk and Titan, over fields of $1^{\prime\prime}\times1^{\prime\prime}$ and $1.4^{\prime\prime}\times1.4^{\prime\prime}$, respectively. In these fusion studies, the integral-field cube samples spectra across the whole reconstructed region. Survey spectroscopy instead places sparse fibers across far larger imaged areas, so a joint model must carry spectral information from the fibers to positions where only images exist.

In computer vision, novel-view synthesis methods fit a scene representation through differentiable rendering, then predict images from new viewpoints. Neural radiance fields represent a scene with a multi-layer perceptron over continuous coordinates \citep{mildenhall2020nerf}, while Gaussian splatting uses explicit Gaussian components whose positions, shapes, and number are adapted during optimization by adaptive density control \citep{kerbl20233dgaussiansplattingrealtime}. Radiance fields with joint image registration \citep{lin2021barf} have also been used to coadd simulated astronomical images \citep{hutton2022barf}. In astronomy, Gaussian sums have long been used to model galaxy light distributions \citep{cappellari2002mge} and to speed up the rendering of standard galaxy profiles \citep{hogg2013mog}. Explicit Gaussians also suit astronomical forward modeling: when the PSF is represented by a Gaussian sum, convolution reduces to combining Gaussian parameters, while affine transformations preserve the Gaussian form when mapping between sky and detector coordinates.

In this work we introduce Spectro-Spatial Survey Fusion, a practical framework that combines heterogeneous astronomical observations by fitting a single field that is continuous in position and wavelength. We represent the field as a sum of anisotropic spatial Gaussians, each carrying a spectrum assembled from a shared dictionary. The component parameters and spectral dictionary are learned jointly, with growth and pruning adapting the Gaussian population to describe source morphology freely without imposing a fixed profile for each object. Each image pixel or spectral bin is predicted through its instrument response, requiring the same spatial and spectral field to account for all the fitted measurements. Once fitted, this field can be queried at different positions and wavelengths or through new instrument responses without refitting.

We demonstrate the framework on an $84^{\prime}\times84^{\prime}$ region of the COSMOS field using 400 Sloan Digital Sky Survey (SDSS) $ugriz$ exposures and 800 fiber spectra. We refer to this joint reconstruction as the `canonical fit'. We first assess how closely the field reproduces the fitted observations, then predict 89 withheld fiber spectra and ten withheld image exposures, without refitting. External imaging from HSC and spectroscopy from DESI provide a generalization test of the field's predictions through instruments that supplied no data to the fit. This framework points toward a common model of the sky that can be refined as observations from additional surveys are incorporated. Jointly fitting observations from several surveys would allow spatial and spectral information resolved by one instrument to constrain what remains uncertain in another, and allow a representation of the sky that contains more information than any single survey could provide.

The paper is organized as follows: Section~\ref{sec:representation} defines the underlying representation and observation operators, and Section~\ref{sec:data_fitting} describes the data and fitting procedure. Sections~\ref{sec:fitted_field} and~\ref{sec:synthesis} present the fitted field and its predictions for new measurements, respectively. Section~\ref{sec:discussion} discusses their implications and limitations, and Section~\ref{sec:conclusions} concludes.

\section{Representation and Operators}
\label{sec:representation}

\subsection{A common measurement model}
\label{sec:measurement_model}

We represent the sky's latent emission by a single, time-independent field. Because the fitted region spans only $84^{\prime}$, we describe sky positions by flat-sky coordinates about the field center $(\alpha_0,\delta_0)$ given in Section~\ref{sec:observations}, $\boldsymbol{x}=\bigl((\alpha-\alpha_0)\cos\delta_0,\,\delta-\delta_0\bigr)$. Let $\mathcal{X}\subset\mathbb{R}^2$ be the fitted region in these coordinates, and let $\Lambda=[2980,11230]$\,\AA\ be the modeled range of observed wavelength. The field
\begin{equation}
    F_\lambda:\mathcal{X}\times\Lambda\to\mathbb{R}_{\geq0}
    \label{eq:field_domain}
\end{equation}
gives the spectral surface brightness at position $\boldsymbol{x}\in\mathcal{X}$ and observed wavelength $\lambda\in\Lambda$ before atmospheric and instrumental blurring, as a flux density per unit wavelength and solid angle. Across the field, the east--west scale of these coordinates differs from the true angular scale by at most 0.06~percent.

Each calibrated image pixel or spectral bin is a datum. We index data by $i\in\mathcal{D}$, where $\mathcal{D}=\mathcal{D}_{\mathrm{img}}\cup\mathcal{D}_{\mathrm{spec}}$ contains all image pixels and spectral bins, and write $e(i)$ for the image exposure or fiber spectrum that contains datum $i$. Images and fiber spectra constrain the same field through their respective observation operators. Let $d_i$ be the calibrated measurement after subtracting the survey-supplied background estimate $B_i$. We write the general-purpose measurement model as

\begin{equation}
    d_i = \widehat{d}_i + \epsilon_i,
    \qquad
    \widehat{d}_i := \mathcal{A}_i[\flam],
    \label{eq:measurement_model}
\end{equation}

where $\mathcal{A}_i$ is the observation operator that maps the field into the instrument-defined units and sampling of datum $i$, $\widehat{d}_i$ is the predicted source flux, and $\epsilon_i$ is measurement noise. The survey background is subtracted from the measurements before fitting and held fixed; it also enters the image noise model (Section~\ref{sec:objective}). Errors in this survey-supplied estimate can bias the inferred field or remain in the residuals if not addressed during data cleaning before fitting (Section~\ref{sec:discussion}). Appendix~\ref{app:notation} summarizes the notation used throughout.

\begin{figure*}
    \centering
    \includegraphics[width=0.9\textwidth]{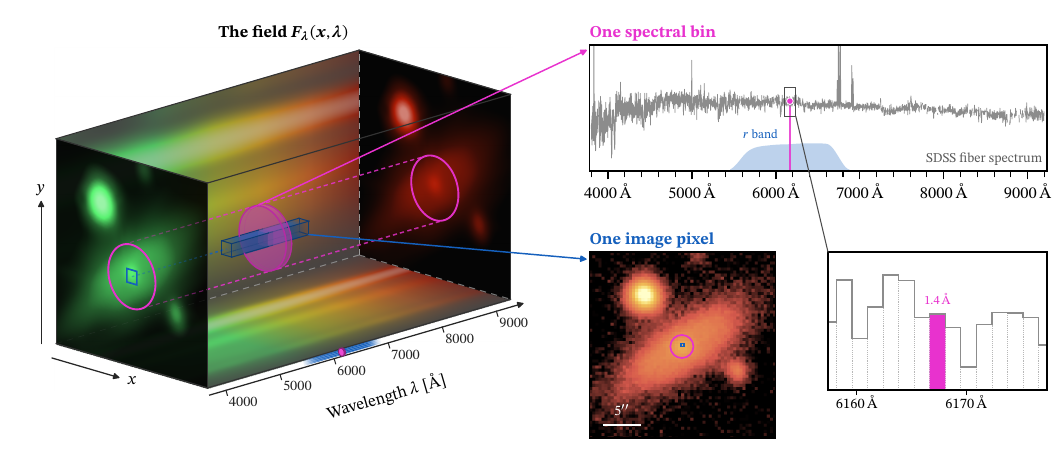}
    \caption{\textbf{A schematic diagram of an SDSS image pixel and a fiber spectral bin for the same galaxy.} The blue rectangular column spans the $r$ band at the position of one pixel. The pink disk spans the fiber aperture in a single $1.4$\,\AA\ bin. Their sizes are not to scale, but emphasize the small spatial footprint and broad-in-wavelength bandpass of an image pixel, compared with the larger spatial aperture and narrow wavelength interval of a spectral bin. The right panels show the two measurements in the observed image and spectrum.}
    \label{fig:measurement_model}
\end{figure*}

Figure~\ref{fig:measurement_model} conceptually illustrates these two kinds of measurements from the same galaxy. Both measurements draw from the same underlying \flam field that is continuous in position \posx and wavelength \lam. An image pixel collects light from a small spatial region across a bandpass that is broad in wavelength, represented by the blue column extending across the wavelength axis. In the same latent field, a spectral bin instead collects light across the fiber aperture with fine spectral resolution but coarse spatial resolution, represented by the pink disk. Together, these measurements constrain the spatial and spectral structure of the same field. The observation operators $\mathcal{A}_i$ include the bandpass weighting, spatial and spectral blurring, and flux calibration needed to predict these measurements.

\subsection{Gaussian components and a shared spectral dictionary}
\label{sec:components}

We represent \flam as a sum of spatial Gaussians, each carrying a spectrum formed from a shared spectral dictionary. Gaussians retain their form under the local affine transformations that are used to map the field's continuous sky coordinates onto each exposure's pixel grid. Convolving two Gaussians adds their covariance matrices, so a Gaussian sum remains a sum of Gaussians after convolution with a Gaussian-mixture PSF, and we evaluate the PSF-convolved components analytically in the image operator (Section~\ref{sec:image_operator}). Fiber-aperture integration, however, still uses numerical quadrature (Section~\ref{sec:fiber_operator}). Here, the components' positions and shapes are fitted across the field, allowing several Gaussians to combine into a source without imposing explicit shape priors or fixed morphological templates.

\begin{figure*}
    \centering
    \includegraphics[width=0.95\textwidth]{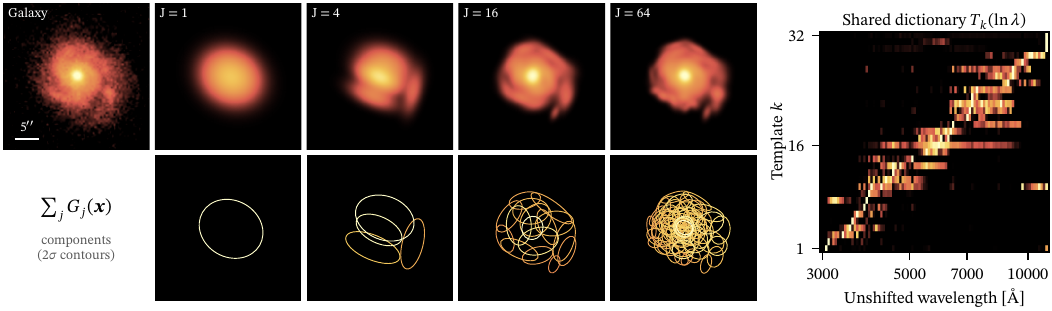}
    \caption{\textbf{Schematic of the spatial and spectral components of the field's underlying representation.} On the left, sums of 1, 4, 16, and 64 Gaussians capture progressively finer structure in an observed galaxy. The ellipses below mark the $2\sigma$ contours of the individual components. The right panel shows the 32 spectral templates learned from the full field, ordered by peak wavelength and each normalized to its own peak. Each component learns its own mixture of these 32 templates.}
    \label{fig:representation}
\end{figure*}

We express the sky's spectral surface brightness as a sum of $J$ Gaussian components:

\begin{equation}
    \flam(\posx,\lam) := \sum_{j=1}^{J} A_j G_j(\posx) s_j(\lam),
    \label{eq:field_components}
\end{equation}

where $A_j\geq0$ is the component amplitude and $s_j:\Lambda\to\mathbb{R}_{\geq0}$ is its spectrum. Each $G_j$ is a two-dimensional Gaussian normalized to unit integral, with center $\boldsymbol{\mu}_j$ and symmetric positive-definite covariance $\boldsymbol{\Sigma}_j$. Two positive standard deviations along its principal axes and a position angle determine its size, elongation, and orientation. The components provide a basis for the light distribution and need not individually correspond to physical objects. The left panels of Figure~\ref{fig:representation} show a conceptual illustration of how an increasing number of Gaussians allows a Gaussian sum to describe more spatial structure. Our field is fitted jointly to images and spectra, with an evolving Gaussian component lifecycle governed by a modified version of the density-control prescription from Gaussian splatting \citep{kerbl20233dgaussiansplattingrealtime}, as described in Section~\ref{sec:optimization}.

To allow for shared information across the field, we represent the component spectra with a shared dictionary of $K$ non-negative templates,

\begin{equation}
    s_j(\lam) := \sum_{k=1}^{K} w_{jk} T_k(\ln\lambda-\zeta_j),
    \label{eq:component_spectra}
\end{equation}

where the coefficients $w_{jk}\geq0$ and shift $\zeta_j$ are specific to component $j$. Each component has one learned shift in log wavelength, applied to all $K$ template contributions, so the same template shapes can be used at different wavelengths. These shifts are free per component and are intended to make the spectral representation more flexible. The $K=32$ templates share a grid of 96 knots spanning 2980--11230\,\AA. The knots are uniformly spaced in log wavelength, and template values are interpolated linearly in wavelength between them. Template values outside the stored interval are set to zero. The amplitudes $A_j$, mixture coefficients $w_{jk}$, and template knot values are kept non-negative through softplus parameterizations. The value of $A_j$ alone does not give the component's flux, as there is a normalization degeneracy among $A_j$, the mixture coefficients, and the templates. We therefore interpret predicted observables and the combined light of components rather than individual normalizations. The right panel of Figure~\ref{fig:representation} shows the dictionary learned by the canonical fit (Section~\ref{sec:what_learned}), with templates normalized to their peaks and ordered by peak wavelength for display. Different component spectra allow the model to represent color variations within a source. Each Gaussian has one position and shape shared across wavelengths, while its brightness varies with wavelength.

The dictionary is learned jointly with the component parameters from the fitted observations and their instrument responses, rather than supplied as a fixed library of physical spectra. Individual templates need not correspond to any physical source type. There is no external pretraining corpus or source catalog that informs the model about where to allocate spectral or spatial capacity. Spectra within fitted fibers can therefore constrain template shapes also used by components elsewhere in the field, while those components retain their own positions, shapes, mixtures, and spectral shifts. This shared basis allows information from spectroscopic measurements to affect predictions away from their original fitted apertures. Section~\ref{sec:withheld_spectra} tests how including spectra in the fit changes predictions for withheld spectra.

\subsection{Rendering the field through an instrument operator}
\label{sec:rendering}

We predict the source light in each pixel or spectral bin by summing the contributions from the field's Gaussian components. We evaluate these contributions at the observed pixel positions and spectral sampling, separating their responses by modality. We also approximate the spatial PSF as independent of wavelength within an image band or an individual fiber spectrum. The predictions below give $\widehat{d}_i=\mathcal{A}_i[\flam]$ in Equation~\ref{eq:measurement_model}. Within the modeled sky region and wavelength range, the same fitted field can predict measurements from other instruments without refitting, provided their response and flux convention are adequately modeled. We demonstrate this with zero-shot HSC imaging and DESI spectroscopy in Section~\ref{sec:new_instruments}.

\subsubsection{Imaging}
\label{sec:image_operator}

For an image datum $i\in\mathcal{D}_{\mathrm{img}}$ in exposure $e(i)$ and band $b(i)$, we first compute the band average $\overline{s}_{bj}$ of each component's spectrum using the system throughput $R_b(\lam)$. We then combine this with the PSF-convolved spatial profile to predict the source surface brightness $\widehat{d}_i$ at the pixel center $\boldsymbol{x}_i$:

\begin{equation}
\begin{aligned}
    \overline{s}_{bj}
        &:= \frac{\int \lambda R_b(\lambda)s_j(\lambda)\,\mathrm{d}\lambda}
                {\int \lambda R_b(\lambda)\,\mathrm{d}\lambda}, \\
    \widehat{d}_i
        &= \sum_j A_j\overline{s}_{b(i)j}
           \bigl(P^{\mathrm{pix}}_i*G_j\bigr)(\boldsymbol{x}_i),
\end{aligned}
\label{eq:image_response}
\end{equation}

where $\boldsymbol{x}_i\in\mathcal{X}$ is the position of the pixel center and $P^{\mathrm{pix}}_i$ is the normalized, pixel-convolved PSF of exposure $e(i)$ at that detector position, expressed in the field's coordinates. Here $*$ denotes convolution in position. The factor of $\lambda$ gives the photon-counting weight. The denominator defines a band-averaged flux density, matching the conversion of the calibrated images to surface-brightness units. Because the PSF is taken to be achromatic within the band, the same spatial response acts on the whole band amplitude. We average each shifted template over the full bandpass, treating its contribution as zero at wavelengths outside the template's range.

Each exposure's survey-supplied world-coordinate system maps its pixel centers to sky coordinates, which we then convert to the flat-sky coordinates of Section~\ref{sec:measurement_model}. Local affine transformations carry the component covariances into detector coordinates, with the Jacobian determinant providing the conversion between pixel area and sky area needed to preserve each Gaussian's total flux. We represent the spatially varying SDSS PSF by fitting signed Gaussian expansions of its supplied eigenimages, combined with the survey's position-dependent coefficients. This gives 40 Gaussian terms at each pixel. The SDSS PSF eigenimages already account for integration over the detector pixels, so we fit this sampled response directly and evaluate the convolved components at pixel centers without integrating over the pixel footprint again.\footnote{The SDSS PSF reconstruction and native sampling are documented at \url{https://www.sdss4.org/dr17/imaging/images/}.} Convolution is analytic for these Gaussian expansions, while the PSF representation and local astrometric linearization remain approximations.

\subsubsection{Fiber spectroscopy}
\label{sec:fiber_operator}

A fiber spectrum combines light across a circular aperture before resolving it in wavelength. For spectrum $e$, let $D_e\subset\mathcal{X}$ be the disk centered on the recorded fiber position, with radius $\theta^{\mathrm{fib}}_e$. The SDSS and Baryon Oscillation Spectroscopic Survey (BOSS) spectrograph fibers \citep{smee2013spectrographs} have diameters of $3^{\prime\prime}$ and $2^{\prime\prime}$, respectively. We convolve each spatial component with a circular Gaussian PSF $P_e$, using the spectroscopic seeing assigned to that spectrum rather than the imaging seeing. Integrating this profile over $D_e$ by numerical quadrature gives $Q_{ej}$, the fraction of component $j$'s light collected by the fiber. Convolving its spectrum with the LSF and averaging over the wavelength interval $\Lambda_i\subset\Lambda$ of spectral bin $i\in\mathcal{D}_{\mathrm{spec}}$ gives $\overline{s}_{ij}$:

\begin{equation}
\begin{aligned}
    Q_{ej} &:= \int_{D_e}(P_e*G_j)(\posx)\,\mathrm{d}^2\boldsymbol{x}, \\
    \overline{s}_{ij} &:= \frac{1}{|\Lambda_i|}
        \int_{\Lambda_i}
        (L_i*s_j)(\lambda)\,\mathrm{d}\lambda,
\end{aligned}
\label{eq:fiber_weights}
\end{equation}

where $|\Lambda_i|$ is the bin width. Here $L_i*s_j$ is the component spectrum's convolution in wavelength with the LSF. We model spectral blurring with a Gaussian LSF $L_i$, normalized so that its integral over wavelength is one. We determine the LSF width from the wavelength-dependent instrumental resolution supplied with each spectrum, treating it as constant within each bin. We approximate the recorded samples as finite-bin averages. The convolution and bin integral can be evaluated analytically for the piecewise-linear component spectra. The LSF and bin widths are of order a few angstroms, much smaller than the template knot spacing, which grows in proportion to wavelength (e.g.\ 84\,\AA\ at 6000\,\AA).

The aperture integral alone is not the flux reported in the SDSS spectra. Their calibration uses standard stars and is normalized to point-source PSF magnitudes.\footnote{See the SDSS spectrophotometric calibration description at \url{https://www.sdss4.org/dr17/algorithms/spectrophotometry/}.} We approximate this convention by dividing the collected source light by the fraction of a centered point source admitted under the same Gaussian seeing. We denote this point-source fraction by $\eta_e$ and use it to normalize the predicted spectral flux density $\widehat{d}_i$ of bin $i$ in spectrum $e=e(i)$:

\begin{equation}
\begin{aligned}
    \eta_e &:= 1-\exp\!\left[-\frac{\bigl(\theta^{\mathrm{fib}}_e\bigr)^2}{2\bigl(\sigma^{\mathrm{see}}_e\bigr)^2}\right], \\
    \widehat{d}_i &= \frac{1}{\eta_e}
        \sum_j A_j Q_{ej} \overline{s}_{ij},
\end{aligned}
\label{eq:fiber_response}
\end{equation}

where $\sigma^{\mathrm{see}}_e$$=\mathrm{FWHM}_e/2.355$ is the standard deviation of the Gaussian seeing profile, and $\mathrm{FWHM}_e$ is its full width at half maximum (FWHM). For a centered point source, dividing by $\eta_e$ recovers its total flux under the assumed PSF model. For an extended source, the same correction does not generally recover all the light missed by the fiber. The predicted spectrum therefore remains dependent on the aperture's size and position through the collected fractions $Q_{ej}$. The approximation uses one wavelength-independent seeing value and a fixed aperture center for each spectrum. It does not reproduce non-Gaussian PSF wings, wavelength-dependent seeing, atmospheric differential refraction, or the full coaddition and calibration procedure.

We retain the native SDSS imaging calibration, which approximates the AB system, and convert the spectra to the same flux scale before fitting. We construct a fixed, wavelength-dependent conversion from the spectroscopic pipeline's five per-band calibration offsets and the filter throughput curves, and apply it consistently to the spectral fluxes, sky estimates, and uncertainties.

\section{Data and Fitting Procedure}
\label{sec:data_fitting}

\subsection{Observations}
\label{sec:observations}

We apply the framework to an $84^{\prime}\times84^{\prime}$ region of the COSMOS field centered at $(\alpha_0,\delta_0)=(150.1192^{\circ},2.2058^{\circ})$ in ICRS, using SDSS imaging and fiber spectroscopy \citep{york2000sdss}. The imaging comprises 82 SDSS fields observed in $ugriz$ bandpasses, giving 410 exposures obtained between May 2000 and February 2001. We use cutouts from the individual exposures at their native sampling of $0.396^{\prime\prime}$ per pixel, retaining each exposure's astrometry and PSF in its observation operator. We use the SDSS-provided frame masks (fpM) to exclude interpolated pixels and those flagged for saturation, cosmic rays, or ghosts from the fit and image statistics.

The spectroscopic sample comprises 889 spectra of galaxies, quasars, and stars observed between December 2000 and March 2011: 557 taken with the SDSS spectrograph \citep{york2000sdss,rockosi2022segue2} and 332 with the BOSS spectrograph \citep{dawson2013boss}. Each SDSS spectrum combines several exposures of the same target. We retain bins with finite fluxes and wavelengths, and positive inverse variance. We exclude a bin if the pipeline flagged it in any of those exposures, even when the other exposures were unflagged. Each retained spectrum must also have imaging coverage.

Figure~\ref{fig:data_coverage} shows the spatial distribution of the fitted and withheld measurements (Section~\ref{sec:held_out_design}). In the left panel, shading counts how many of the 400 fitted imaging exposures cover each position, summed over all five bands; darker regions have more overlapping coverage. The red dashed outlines mark the footprints of the ten withheld exposures, with labels identifying the withheld band or bands. Fitted images in other bands, and same-band overlaps from other runs or camera columns, still constrain the sky inside these outlines. In the right panel, filled gray symbols locate the 800 fitted spectra and open red symbols the 89 withheld spectra. Circles, squares, and triangles distinguish galaxies, quasars, and stars, respectively.

We fit the field directly to the measured image and spectral fluxes, without using a pre-existing object catalog to fix the source models. Components are seeded from peaks detected in the fitted images (Section~\ref{sec:optimization}). Section~\ref{sec:new_instruments} presents zero-shot experiments with HSC PDR3 imaging \citep{aihara2022hscpdr3} and DESI DR1 spectra \citep{desicollaboration2026dr1} of the same region, and neither dataset contributes to the initialization or fitting of the field and its shared spectral dictionary. Part of the SDSS data is withheld from the fitting objective, as described in Section~\ref{sec:held_out_design}.

\subsection{Fitted and withheld measurements}
\label{sec:held_out_design}

\begin{figure*}
    \centering
    \includegraphics[width=0.9\textwidth]{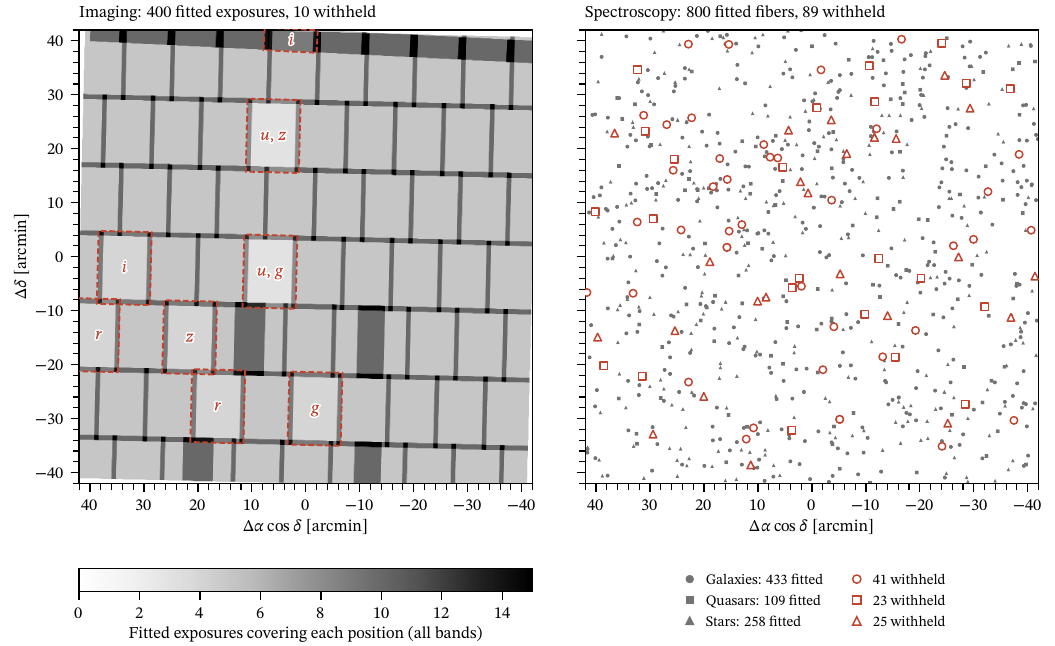}
    \caption{\textbf{SDSS imaging and spectroscopy coverage across the $84^{\prime}\times84^{\prime}$ COSMOS field.} The shading on the left shows the total number of fitted $ugriz$ exposures covering each position. Dashed red outlines mark the ten withheld exposures, labeled by band. On the right, filled gray symbols mark the 800 fitted spectra and open red symbols the 89 withheld spectra. Circles mark galaxies, squares mark quasars, and triangles mark stars.}
    \label{fig:data_coverage}
\end{figure*}

We assess predictions of withheld SDSS measurements using the fixed split shown in red in Figure~\ref{fig:data_coverage}. Ten complete image exposures, two per band, and 89 fiber spectra are omitted from the fitting objective. The remaining 400 exposures and 800 spectra constrain the field. The fitted spectra comprise 433 galaxies, 109 quasars, and 258 stars, while the withheld spectra comprise 41 galaxies, 23 quasars, and 25 stars. Spectral observations were assigned to proximity groups, linking target positions separated by less than $10^{\prime\prime}$, before withholding whole groups.

The information available within a withheld image depends on the coverage of the fitted exposures. Consecutive SDSS fields share 128 rows of the same detector readout. When both copies of a shared detector sample would enter the fit, we retain only one. If either copy is withheld, through either a whole-exposure exclusion or the random pixel selection described below, both copies are excluded from the fitting objective. Thus no withheld detector sample enters the objective through a neighboring field. Overlapping observations from different runs or camera columns are retained as separate measurements. For nine of the ten withheld exposures, 86--87~percent of the valid pixels have no fitted same-band coverage after the overlap exclusions and frame masking. For the tenth, the fraction is 12~percent, because most of its area within the field is also imaged by fitted exposures from another run. We evaluate the different coverage regions separately (Section~\ref{sec:withheld_exposure}).

In addition to the whole-observation exclusions, two percent of individual image pixels and spectral bins are omitted from the objective, serving as development diagnostics. The image-peak initialization uses the full source images of the fitted exposures, including pixels later excluded from the objective at random, by overlap deduplication, or by the SDSS frame masks, so these randomly omitted image pixels are not an independent test. The withheld spectra and exposures are rendered through their respective operators and presented in Section~\ref{sec:withheld} as a generalization test.

We assess both fitted and withheld observations by measuring the residuals relative to uncertainties derived from the observed data. This puts discrepancies in image pixels and spectral bins on a common noise scale. For each datum $i$, we compare the background-subtracted measurement $d_i$ with its prediction $\widehat{d}_i$ (Equation~\ref{eq:measurement_model}), and let $\sigma_i$ be its observational uncertainty. For image pixels, we calculate $\sigma_i$ from the observed source-plus-sky counts and the survey's gain, read-noise, and dark-current parameters. For spectral bins, we derive $\sigma_i$ from the pipeline inverse variances and rescale it with the fluxes as described in Section~\ref{sec:fiber_operator}. These same uncertainties define the source-pixel selection below. We define the pull $\chi_i$ of each datum and the mean squared pull (MSP) of a set of data $\mathcal{C}\subseteq\mathcal{D}$ as
\begin{equation}
    \chi_i := \frac{d_i-\widehat{d}_i}{\sigma_i},
    \qquad
    \mathrm{MSP}(\mathcal{C}) := \frac{1}{|\mathcal{C}|}\sum_{i\in\mathcal{C}}\chi_i^2,
    \label{eq:pull_msp}
\end{equation}
where $\mathcal{C}$ is, for example, the source-selected pixels of one exposure or the bins of one spectrum. The MSP is the chi-square per datum of that set.

In the images, the large number of pixels with little source light can dilute the residuals around bright sources in an average over the whole exposure. We therefore report statistics separately for source-selected pixels, with $d_i>5\sigma_i$, and for the remaining pixels, which we call sky pixels even though they may contain faint source light. We calculate these statistics on the original pixels of each exposure, before resampling the images for display. As in the fit, valid pixels have finite fluxes and uncertainties with $\sigma_i>0$ and pass the SDSS frame-mask selection in Section~\ref{sec:observations}. For fitted images, we pool pixels within the stated band or region using the same exclusions as the objective, including duplicate samples and the randomly excluded pixels. For withheld images, we use valid pixels within each stated coverage region.

For the spectra, we compute statistics for each observation using all bins retained by the selection in Section~\ref{sec:observations}, including the randomly excluded bins within fitted spectra. Medians over these per-spectrum statistics summarize the population with each spectrum given equal weight. We also report a pooled MSP over all retained bins, in which spectra with more bins contribute more to the result.

A residual that is small compared with the source flux can still be large compared with the measurement uncertainty. We therefore also report predicted-to-observed flux ratios, summing over the stated image pixels or integrating over retained spectral bins with their individual wavelength widths. Masked wavelength gaps contribute no flux to these integrals. Agreement in summed flux can hide an excess in one pixel that cancels a deficit in another. For source-selected image pixels, we also measure the fractional absolute residual, $\sum_i|d_i-\widehat{d}_i|/\sum_i d_i$, to quantify the discrepancy without this cancellation.

\subsection{The objective}
\label{sec:objective}

We combine the fitted image pixels and spectral bins in a common objective, weighting their squared residuals inversely by their noise variances. Let $\boldsymbol{\theta}$ denote the component parameters and shared spectral dictionary, and let $\mathcal{D}^{\mathrm{fit}}_{\mathrm{img}}\subset\mathcal{D}_{\mathrm{img}}$ and $\mathcal{D}^{\mathrm{fit}}_{\mathrm{spec}}\subset\mathcal{D}_{\mathrm{spec}}$ contain the image pixels and spectral bins that enter the fitting objective after the exclusions in Section~\ref{sec:held_out_design}. Using $d_i$ and the predictions $\widehat{d}_i(\boldsymbol{\theta})$, the parameter-dependent part of the objective at update $t$ is
\begin{equation}
\begin{aligned}
    \mathcal{L}_t(\boldsymbol{\theta}) = \frac{1}{2N_{\mathrm{ref}}}\Bigg[&
        \sum_{i\in\mathcal{D}^{\mathrm{fit}}_{\mathrm{img}}}\frac{[d_i-\widehat{d}_i(\boldsymbol{\theta})]^2}{v_i^{(t)}} \\
        &+ \gamma_{\mathrm{spec}}\sum_{i\in\mathcal{D}^{\mathrm{fit}}_{\mathrm{spec}}}
        \frac{[d_i-\widehat{d}_i(\boldsymbol{\theta})]^2}{\sigma_i^2}\Bigg],
\end{aligned}
\label{eq:fitting_objective}
\end{equation}

where $v_i^{(t)}$ is the image variance used for that update and $\sigma_i^2$ is the spectral variance derived from the pipeline inverse variance and converted with the flux. The spectral weight $\gamma_{\mathrm{spec}}$ is a fixed hyperparameter controlling the relative contribution of spectroscopy to the joint fit. We adopt $\gamma_{\mathrm{spec}}=5$ for the canonical fit. Each fitted pixel or bin contributes one term. The fixed normalization constant $N_{\mathrm{ref}}=875{,}100{,}989$, equal to the total number of fitted image pixels and spectral bins, sets the overall scale without changing the relative weights. With the variances held fixed, Equation~\ref{eq:fitting_objective} is the weighted sum of independent Gaussian negative log-likelihood terms, apart from normalization terms that contribute no parameter gradient. We add no explicit prior or regularization term, either on the field or on the shared dictionary. The Gaussian parameterization, non-negativity, finite dictionary, initialization, and density control nevertheless act as implicit structural assumptions.

For image pixels, we use the standard CCD variance model, combining source and sky shot noise with read noise and dark-current noise. Calculating the variance directly from the measured counts gives greater weight to measurements that fluctuate below their expected values, which can bias the fitted flux downward. We instead calculate the shot-noise variance from the current predicted source-plus-sky signal. At the current parameters $\boldsymbol{\theta}_t$, the variance in calibrated flux units is
\begin{equation}
    v_{i,\mathrm{model}}^{(t)} =
    \underbrace{\frac{\max\!\left[\widehat{d}_i(\boldsymbol{\theta}_t)+B_i,\,0\right]}{g_i}}_{\substack{\text{source and sky}\\\text{shot noise}}}
    +\underbrace{\bigl(\sigma^{\mathrm{det}}_i\bigr)^2}_{\substack{\text{read noise and}\\\text{dark current}}},
    \label{eq:fitting_image_variance}
\end{equation}
where $B_i$ is the fixed background estimate. The effective gain $g_i$ converts predicted flux to counts, combining the survey gain with the flux calibration, and $\sigma^{\mathrm{det}}_i$ is the detector noise from read noise and dark current, expressed in flux units. Both quantities follow the detector-column calibration of the exposure. We use $v_i^{(t)}=v_{i,\mathrm{model}}^{(t)}$ unless $|d_i-\widehat{d}_i(\boldsymbol{\theta}_t)|>10\sigma_i$, in which case we retain the data-derived variance $\sigma_i^2$ from Section~\ref{sec:held_out_design}. This guard prevents an underestimated model variance from giving excessive weight to a poorly fitted pixel. The image variances and the guard decision are held fixed when taking parameter gradients and recomputed at the next update. For pixels within the guard and with positive source-plus-sky predictions, holding the variance fixed gives the same gradient as a shifted Poisson negative log-likelihood, with the read-noise and dark-current variance represented by equivalent counts. The image pulls and MSPs reported in the paper continue to use the data-derived uncertainties defined in Section~\ref{sec:held_out_design}, so that different predictions are evaluated against the same noise estimates.

\subsection{Optimization and adaptive density control}
\label{sec:optimization}

\begin{figure*}
    \centering
    \includegraphics[width=0.85\textwidth]{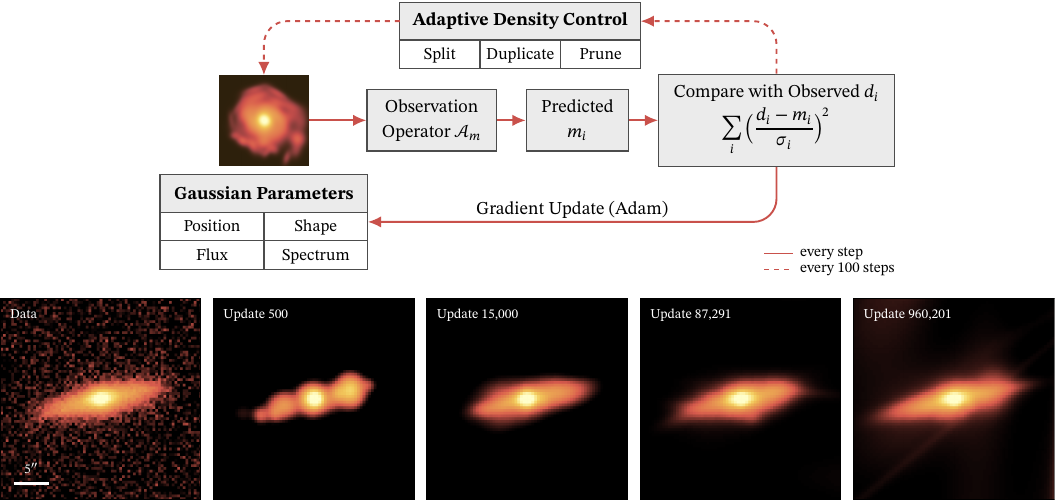}
    \caption{\textbf{Overview of the field's fitting procedure.} At each step, the field is rendered through the observation operator to predict image pixels and spectral bins. These predictions are compared with the data using the loss defined in Section~\ref{sec:objective}, and Adam uses gradients of the fitting error to update the model parameters. Solid arrows show this cycle occurring every step, while dashed arrows show adaptive density control, which can split, duplicate, or prune components every 100 steps. The bottom row shows an observed $r$-band galaxy's evolving representation at four stages of the fit, ending after 11 epochs.}
    \label{fig:fitting_loop}
\end{figure*}

We initialize $10^5$ Gaussian components from peaks in the fitted source images, then alternate parameter updates with changes to the component population, as shown in the schematic diagram of Figure~\ref{fig:fitting_loop}. We search the full source images of all fitted exposures (Section~\ref{sec:held_out_design}) to distribute the initial components near observed emission throughout the field. We merge repeated detections across bands and overlapping exposures so that the same brightness peak does not seed multiple sets of components. To represent both compact and extended emission from the outset, we initialize Gaussians of several widths around each retained peak. Their initial amplitudes are set from the measured peak brightness. The shared templates begin as distinct smooth slopes spanning blue to red, giving the initial dictionary a range of broad spectral shapes.

Each update uses a batch of $10^4$ image pixels and 64 spectral bins to form an unbiased estimate of the full-data objective. We draw pixels across the images used in the fit, with each image contributing to the batches in proportion to its number of included pixels. Within each image, we favor pixels with larger observed source signals relative to their noise, since most pixels contain little source light. A five-percent uniform contribution keeps every pixel included in the fit eligible, and the sampling probabilities remain fixed throughout the fit. We weight each sampled image term in inverse proportion to its selection probability, compensating for the extra visits to pixels containing source light so that their expected contribution to the objective is unchanged. Spectral bins are drawn from shuffled passes through the fitted-bin list. We multiply the corrected image average and the spectral average by their respective fitted populations, then apply the spectral weight and normalization in Equation~\ref{eq:fitting_objective}.

We minimize this sampled objective using Adam \citep{kingma2015adam}. A sparse variant of Adam updates only the components that contribute to the sampled pixels and bins, while the shared spectral templates are updated at every step. We run the canonical fit for 11 image epochs, which takes 8~h on a single NVIDIA B200 GPU, with a peak allocated GPU memory of 11~GB. One image epoch corresponds to drawing approximately as many pixels as are included in the fit, although an individual pixel may be sampled more than once or not at all. The optimizer settings and implementation details are given in Appendix~\ref{app:optimization_settings}.

To add spatial freedom during fitting, we adapt the gradient-guided cloning and splitting scheme of \citet{kerbl20233dgaussiansplattingrealtime}, using a local port of gsplat's densification strategy \citep{ye2025gsplat}. The positional gradient measures how the objective responds to moving a component. Components whose mean positional-gradient magnitude exceeds a threshold are considered for duplication or splitting. Smaller candidates are duplicated at the same position and size, so the two components can then move and change shape separately. Larger candidates are split into two narrower Gaussians at positions drawn from the parent, so they can fit smaller features. Both operations preserve the parent's spectral shape and divide its amplitude equally between the resulting pair. This conserves total flux at each wavelength in our additive field. Because gradient magnitudes depend on the fitting objective, we calibrate the growth threshold from the run's own gradients. We use the median PSF width of the fitted images to set the size threshold between duplication and splitting. The quantitative thresholds and refinement schedule are given in Section~\ref{app:component_lifecycle}.

Pruning removes components whose estimated PSF-convolved peak is negligible relative to the image noise in every band, together with components above a maximum width. For pruning, we estimate each component's peak brightness after blurring it with a circular Gaussian PSF, using the median PSF width in each band. Image predictions retain the full observation operator described in Section~\ref{sec:image_operator}. After growth stops, we continue updating and pruning the components until the fitting budget is reached.

\section{The Fitted Field}
\label{sec:fitted_field}

\subsection{Fitted images across the field}
\label{sec:across_field}

\begin{figure*}
    \centering
    \includegraphics[width=0.9\textwidth]{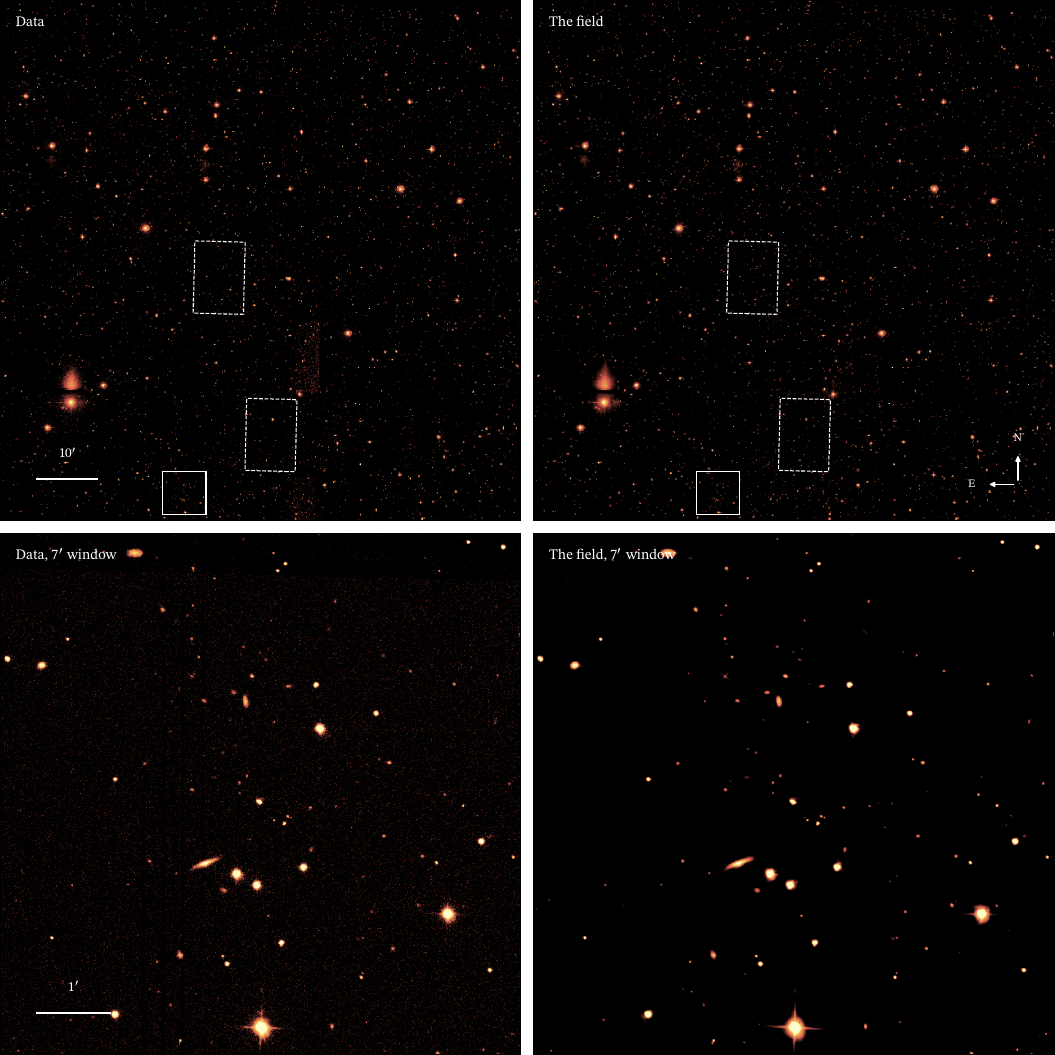}
    \caption{\textbf{Observed SDSS $g$-band imaging and the field's predictions across the $84^{\prime}\times84^{\prime}$ COSMOS field.} The field is represented by 2.86 million Gaussian components fitted to 873 million $ugriz$ image pixels and 800 fiber spectra. The top row compares mosaics of the observations (left) and predictions (right). Dashed outlines mark regions covered only by withheld $g$-band exposures. The bottom row enlarges the $7^{\prime}\times7^{\prime}$ region marked by the solid square, showing galaxies and stars at the native SDSS pixel scale of $0.396^{\prime\prime}$.}
    \label{fig:whole_field}
\end{figure*}

The joint fit reconstructs an $84^{\prime}\times84^{\prime}$ patch of the COSMOS sky as a field continuous in position and wavelength, and closely reproduces the SDSS images and fiber spectra that entered the fit. Figure~\ref{fig:whole_field} compares the observed SDSS $g$-band mosaic with the corresponding mosaic rendered from the fitted field. The full mosaic includes the withheld exposures, with dashed outlines marking the regions covered only by withheld $g$ images. We examine the withheld exposures separately in Section~\ref{sec:withheld_exposure}.

The $7^{\prime}$ inset window, marked by the solid white square, shows how the same field represents various astrophysical objects such as the elongated disk of an edge-on galaxy, faint neighboring galaxies, and stars. These morphologies emerge from the joint fit without a fixed parametric profile or explicit morphological prior for any source. The enlargement in the figure uses the native SDSS pixel scale of $0.396^{\prime\prime}$, with the same stretch used for the data and predictions.

We quantify the fit across all 400 fitted exposures using the pixel selections and data-derived uncertainties defined in Section~\ref{sec:held_out_design}. Across the five bands, the sky-pixel residuals are comparable in magnitude to the observational noise, with MSP values of 0.98--1.06. Their mean signed pull measures the average offset: its values of $-0.06$ to $-0.02$ indicate that the predictions lie slightly above the data in every band. For the source-selected pixels, the summed predicted flux is only 0.4--1.7~percent below the observed flux across the five bands. The fractional absolute residual is 5.4--6.8~percent, indicating larger differences in the distribution of light that partially cancel when the flux is summed.

\subsection{Images and spectra of the same objects}
\label{sec:both_modalities}

\begin{figure*}
    \centering
    \includegraphics[width=.9\textwidth]{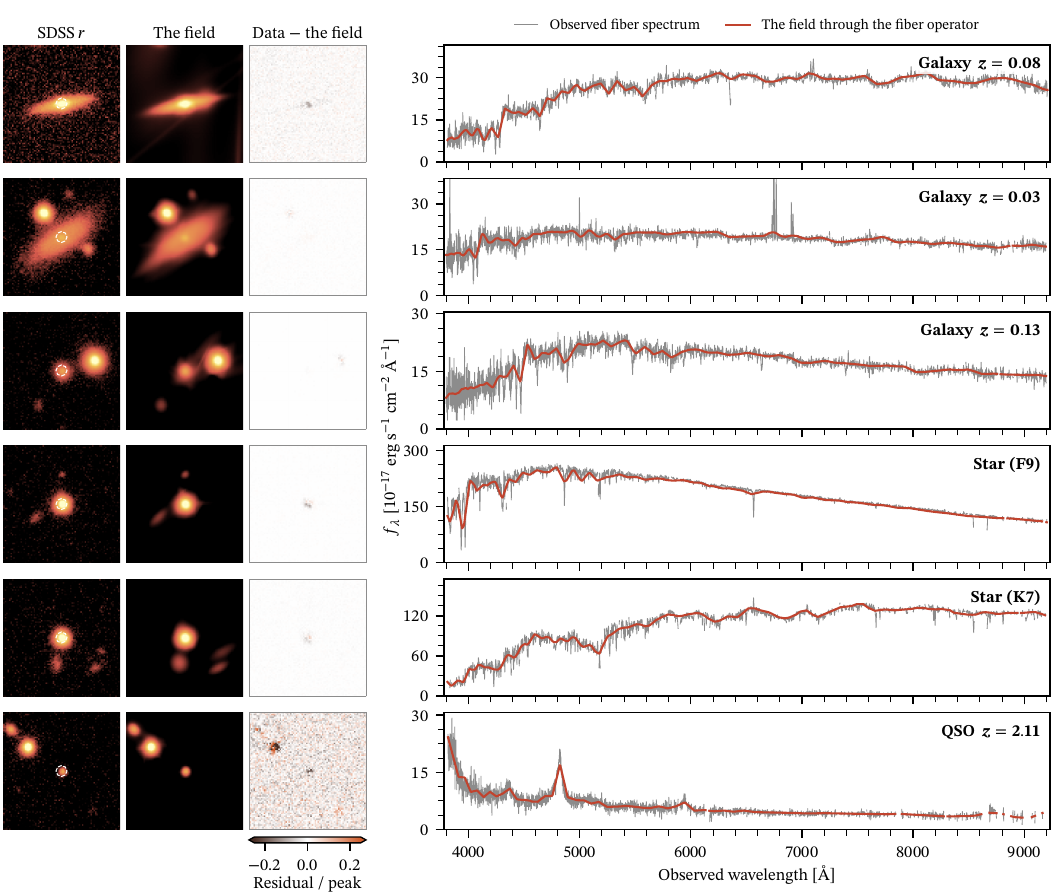}
    \caption{\textbf{Joint fit of the field to images and spectra of three galaxies, two stars, and a quasar.} The three image columns show the observed SDSS $r$-band image, the prediction, and their residual. Dashed circles mark the fiber apertures used for the spectra shown on the right, where gray curves show the observations and red curves show the field's predictions through the same fibers. A single fitted field produces all the image and spectral predictions shown here, through their respective observation operators.}
    \label{fig:fitted_objects}
\end{figure*}

The agreement across the images extends to the spectra of the sources within them. Because the same components carry both the spatial profile and the spectrum, the field captures the extended profiles of galaxies and the compact images of stars and quasars, together with their distinct spectral continua and broad emission features, all built from the same spectral dictionary. Figure~\ref{fig:fitted_objects} shows the spatial and spectral views together for six selected objects: three galaxies, two stars, and a quasar. Each row shows a cutout from a fitted SDSS $r$-band exposure, its rendered image counterpart, the observed spectrum at the marked fiber position, and the field's spectral prediction. Across these examples, the field closely reproduces both the distribution of light in the images and the broad structure of the spectra. This agreement spans objects with very different morphologies and spectral shapes, demonstrating the flexibility of the joint representation across galaxies, stars, and quasars.

The close agreement seen in the selected examples extends to the wider fitted sample, with 96~percent of the 800 spectra having integrated fluxes within ten percent of the observations. Table~\ref{tab:fitted_spectra} shows that integrated-flux agreement persists across the three brightness terciles, with median flux ratios of 0.993, 0.993, and 0.977 for the brightest, middle, and faintest terciles, respectively. The fit can match the total flux of a spectrum without reproducing every feature accurately, and small mismatches stand out more clearly in the brighter spectra because their relative uncertainties are smaller. The median MSP is 0.99 in the faintest tercile and 4.37 in the brightest.

\begin{table}
    \centering
    \caption{Integrated-flux ratios and residuals for the 800 fitted spectra, by brightness tercile. Spectra are ranked by their observed source flux integrated over the bins that pass quality cuts. $N$ is the number of spectra, and the signal-to-noise ratio (S/N) is the median over spectra of each spectrum's median flux-to-uncertainty ratio per bin. The flux ratio is the predicted-to-observed integrated flux (Section~\ref{sec:held_out_design}), given as its median, its interquartile range (IQR), and the percentage of spectra within ten percent of unity. MSP is reported as the median of the tercile.}
    \label{tab:fitted_spectra}
    % generated by output/fitted_spectra_table_20261019/make_fitted_spectra_table.py; do not hand-edit
\begin{tabular}{@{}lrrrcrr@{}}
\hline
Tercile & $N$ & S/N & \multicolumn{2}{c}{Flux ratio} & Within & MSP \\
 & & & median & $[p_{25},\,p_{75}]$ & 10\% & \\
\hline
Brightest & 266 & 32.5 & 0.993 & $[0.984,\,1.002]$ & 97.4\% & 4.37 \\
Middle & 266 & 9.2 & 0.993 & $[0.974,\,1.007]$ & 96.2\% & 1.44 \\
Faintest & 268 & 1.9 & 0.977 & $[0.957,\,1.003]$ & 94.8\% & 0.99 \\
\hline
All & 800 & 9.3 & 0.989 & $[0.973,\,1.004]$ & 96.1\% & 1.45 \\
\hline
\end{tabular}

\end{table}

In Figure~\ref{fig:fitted_objects}, the field follows the F9 star's broad continuum peak and decline toward longer wavelengths, as well as the K7 star's rising continuum. Their integrated fluxes agree with the observations to within 2.1~percent, even though both lie among the ten percent of fitted stars with the highest MSPs. Several narrow absorption troughs remain too shallow. The $z=0.03$ galaxy introduced in Figure~\ref{fig:measurement_model} shows a similar pattern, with a closely matched continuum and narrow emission peaks that remain underpredicted. Section~\ref{sec:discussion} considers the template knot spacing relative to the widths of these features.

\subsection{The fitted components and spectral dictionary}
\label{sec:what_learned}

Across the fitted field, 2.86 million Gaussian components draw on the same 32 spectral templates. The bottom row of Figure~\ref{fig:fitting_loop} follows a single galaxy through the field's growth from 100,000 to 5.64 million components and subsequent pruning to this final population, showing how the representation is refined across updates. Most Gaussian components are narrower than the seeing and combine with broader components to describe the compact and extended emission across the field. The observations constrain their combined light through the instrument responses, so their individual shapes need not trace resolved physical structure, as discussed in Section~\ref{sec:discussion}.

From the smooth slopes used at initialization, the spectral dictionary developed overlapping shapes with peaks at different wavelengths, shown in the right panel of Figure~\ref{fig:representation}. Their wavelength extents are broad, with a median of approximately 1000\,\AA\ enclosing the central 68~percent of template strength on the log-wavelength grid, while variation from knot to knot remains visible within them. These learned shapes provide a common set of contributions that the fit combines in different proportions across the field, but should be interpreted as shared spectral building blocks rather than spectra of individual objects. An object's spectrum can combine light from many Gaussian components, each carrying its own mixture of these templates.

The fitted component spectra combine several of these shapes, with a median effective count equivalent to about six equally contributing templates. This effective count is unchanged by the normalization rescalings in Section~\ref{sec:components}. The contrasting galaxy, stellar, and quasar spectra in Figure~\ref{fig:fitted_objects} are assembled from these mixtures and shifts, summed over the multiple Gaussians contributing to each fiber. Their close continuum agreement is therefore achieved with a common dictionary learned jointly from the 800 fitted spectra and the imaging. The joint fit closely reproduces the spatial structure and broad spectral diversity of the fitted sources within one continuous sky field, without prescribing a separate model for each object. Section~\ref{sec:synthesis} tests what this fitted representation predicts for measurements excluded from the fit and, zero-shot, for observations from other instruments.

\section{Novel Observation Synthesis}
\label{sec:synthesis}

\subsection{Withheld observations}
\label{sec:withheld}

We first predict the 89 SDSS/BOSS fiber spectra and ten SDSS image exposures omitted from the fitting objective (Section~\ref{sec:held_out_design}). The field and its shared dictionary remain fixed, so these comparisons test how the fitted measurements constrain predictions for the withheld spectra and images.

\subsubsection{Withheld spectra and what spectroscopy contributes}
\label{sec:withheld_spectra}

Broadband images constrain integrated fluxes but leave finer spectral structure uncertain. The spectra fitted elsewhere in the field can inform that structure through the shared dictionary. We test their contribution by comparing the joint fit with an images-only control at the 89 withheld fiber positions, with both predictions made without refitting. The control uses the same imaging, random seed, recipe, and update budget, with all spectra excluded and its objective normalized by the fitted pixel count alone.

\begin{figure*}
    \centering
    \includegraphics[width=0.8\textwidth]{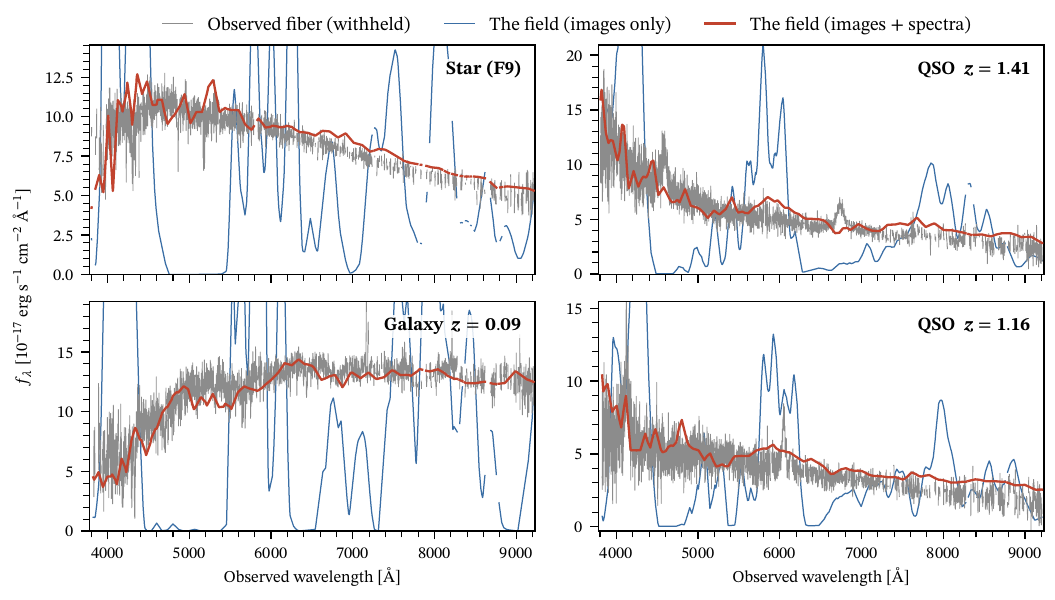}
    \caption{\textbf{Predicting spectra of withheld objects.} Predictions of four withheld SDSS fiber spectra, showing a star, a galaxy, and two quasars. Gray curves show the observed spectra, which never entered the fit during training. Blue curves show predictions from the field when fitted to images alone, while red curves show predictions from the canonical fit of the field, which uses the same images and 800 spectra elsewhere in the field. The joint fit follows the observed continua more closely, generating spectra for objects constrained locally by broadband imaging. In these predictions, spectroscopy elsewhere constrains the spectral dictionary, improving predictions of withheld spectra beyond the fitted fiber positions.}
    \label{fig:withheld_spectra}
\end{figure*}

Figure~\ref{fig:withheld_spectra} illustrates the improvement for selected star, galaxy, and quasar spectra: the joint predictions follow the observed continua much more closely, while the images-only predictions show large excursions with wavelength. Across all 89 withheld spectra, the joint fit has the lower MSP for 88, with a median MSP of 6.01 compared with 98.5 for the control. Repeating both fits with two further random seeds gives the same result, with the joint fit having the lower MSP for 88 of the 89 spectra in each pair. Including spectroscopy therefore substantially improves spectral predictions away from the fitted fiber positions across all three seeds of this field.

For a typical withheld spectrum, the joint prediction still has residuals larger than the observational uncertainties. The median MSP of 6.01 for the withheld sample compares with 1.45 for the 800 fitted spectra in Table~\ref{tab:fitted_spectra}. The median predicted-to-observed integrated-flux ratio is 0.99 for the 55 SDSS-spectrograph spectra observed through $3^{\prime\prime}$ fibers and 0.90 for the 34 BOSS spectra observed through $2^{\prime\prime}$ fibers. Thus the joint fit can follow the broad spectral shapes at withheld positions while retaining appreciable errors in both spectral structure and flux. The F9 star's narrow absorption features in Figure~\ref{fig:withheld_spectra} are not fully reproduced. Section~\ref{sec:discussion} considers the spectral parameterization and its limitations.

\subsubsection{\texorpdfstring{Withheld exposures}{Withheld exposures}}
\label{sec:withheld_exposure}

\begin{figure*}
    \centering
    \includegraphics[width=0.9\textwidth]{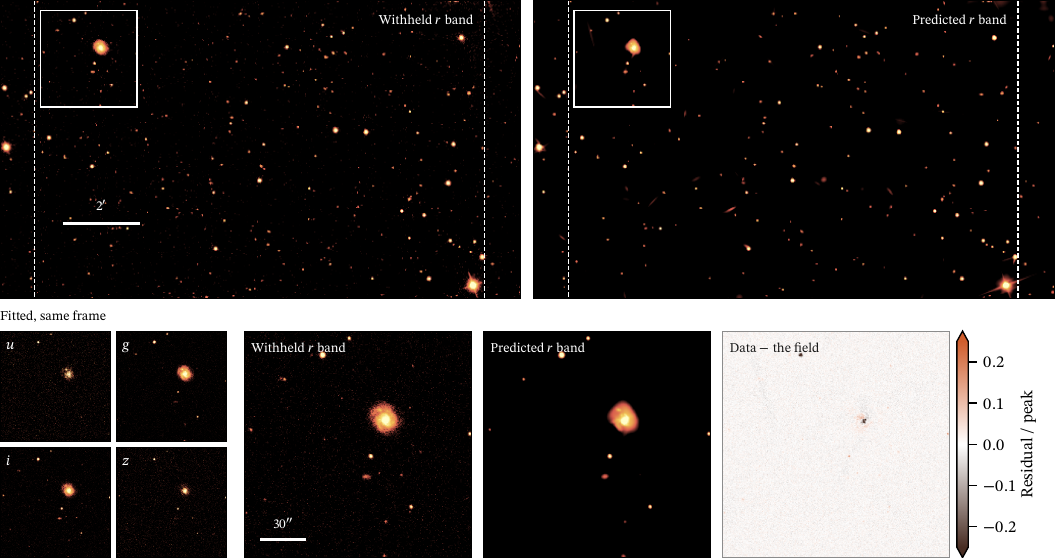}
    \caption{\textbf{Prediction of an SDSS $r$-band exposure withheld from the fit.} The top row compares the observation (left) with the field's prediction (right). Dashed lines enclose the region with no fitted $r$-band coverage, while the edge strips retain $r$-band measurements from another run. The bottom row enlarges the region marked by the solid square, showing the fitted $u$, $g$, $i$, and $z$ images alongside the withheld $r$-band image, its prediction, and their residual. Each Gaussian carries a continuous spectrum built from the shared dictionary and constrained by the other fitted bands, allowing the field to render the $r$-band image even where no $r$-band imaging entered the fit.}
    \label{fig:withheld_exposure}
\end{figure*}

We withhold ten complete image exposures from the fit, two in each of the five SDSS bands. Each withheld exposure removes all measurements from one image in a single band, while the remaining fitted images and spectra continue to constrain the same sky. Figure~\ref{fig:withheld_exposure} shows an $r$-band example for which the corresponding $u$, $g$, $i$, and $z$ exposures were all fitted. After the overlap exclusions and masking, 86~percent of the scored pixels have no fitted $r$-band coverage. The two edge strips retain fitted $r$-band measurements from another SDSS run, and we assess them separately. Copies of withheld detector samples in neighboring fields are also excluded from the objective, as described in Section~\ref{sec:held_out_design}.

The predicted $r$-band image reproduces the broad spatial structure of the withheld observation, including the extended emission around the galaxy in the zoom. To measure agreement around sources, we apply the $d_i>5\sigma_i$ pixel selection from Section~\ref{sec:held_out_design} separately within each coverage region. The summed predicted flux over these pixels is 9.9~percent below the observation in the region without fitted $r$-band coverage and 9.3~percent above in the covered edge strips. The corresponding sky-pixel MSPs are 1.06 and 1.08. Across all ten withheld exposures, source-selected pixels in regions without fitted same-band coverage instead have MSPs of 28--603 and summed predicted-to-observed flux ratios of 0.45--1.16. The largest flux deficits occur in the two $u$ exposures, with ratios of 0.64 and 0.45. For both, the corresponding $g$ or $z$ exposure was also withheld. These withheld exposures show that the field can reproduce broad spatial structure in an excluded band, but the summed flux in source-selected pixels can be substantially biased where no fitted same-band imaging is available.

\subsection{New instruments and operators}
\label{sec:new_instruments}

The fitted spectral surface-brightness field can be queried through different bandpasses, angular resolutions, and apertures to predict observations from other instruments. We test this by virtually reconstructing the instrument operators of HSC PDR3 imaging \citep{aihara2022hscpdr3} and DESI DR1 spectra \citep{desicollaboration2026dr1}, then using them to synthesize the corresponding images and spectra. These predictions are zero-shot: the field and its shared dictionary are learned from SDSS observations alone, with no HSC or DESI measurements used in the fit.

\subsubsection{A new instrument: HSC imaging}
\label{sec:hsc}

\begin{figure*}
    \centering
    \includegraphics[width=0.85\textwidth]{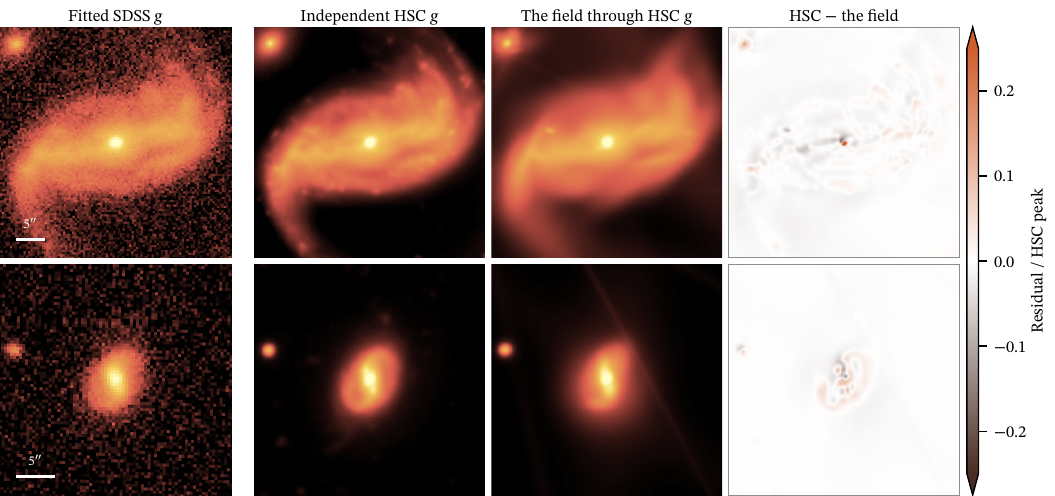}
    \caption{\textbf{Zero-shot HSC imaging.} We render the field fitted to SDSS alone through the HSC $g$-band operator, using the HSC bandpass, PSF, and pixel grid. Each row shows a spiral galaxy. The first column shows the fitted SDSS $g$-band image and the second shows the independent HSC $g$-band observation that was not in training. The third column shows the field's prediction through the reconstructed HSC imaging operator, and the fourth shows the HSC-minus-field residual. Neither galaxy has a fitted SDSS fiber within $10^{\prime\prime}$ of its center.}
    \label{fig:hsc_operator}
\end{figure*}

We first test operator transfer by querying the SDSS-fitted field to predict HSC $g$-band images. Figure~\ref{fig:hsc_operator} compares the observed and predicted HSC images of two selected regions, an extended barred spiral galaxy in the top row and a more compact barred spiral galaxy in the bottom row, alongside the SDSS cutouts from those locations. For these two selected galaxies, the predictions through the virtual HSC instrument response reproduce the extended light distribution in the top row and the broad pattern of the more compact spiral in the bottom row. To make these predictions, we render the field through the HSC $g$ bandpass and local PSF on each coadd's own pixel grid, using the HSC-provided astrometry and AB calibration.\footnote{As for SDSS, we approximate the PSF with a sum of Gaussians. For HSC, we fit the supplied local PSF image directly with non-negative weights and explicitly integrate the predicted light over each pixel.}

Predicted-to-observed HSC flux ratios are measured within fixed $6^{\prime\prime}$-radius apertures centered on the DESI DR1 catalog positions of the selected galaxies (Section~\ref{sec:desi}). SDSS adapts the surface-brightness quantity of \citet{petrosian1976} and sets the flux aperture radius to twice the $r$-band Petrosian radius to include most of a galaxy's light while limiting sky noise \citep{strauss2002}. Of the 80 galaxies without a fitted SDSS spectrum within $10^{\prime\prime}$, 79 have a unique match in the SDSS DR17 catalog \citep{abdurrouf2022sdssdr17}. Their median Petrosian aperture radius is $6.06^{\prime\prime}$, so we use a fixed $6^{\prime\prime}$ radius for the HSC flux ratios. The observed and predicted aperture fluxes are measured at HSC's native resolution using the same unmasked pixels. The field remains fixed, and the predictions are not rescaled to match the observed HSC fluxes. The HSC mask, coverage, and PSF-quality cuts retain 49 of the 80 galaxies without a fitted SDSS spectrum. These galaxies have a median flux ratio of 1.01 and an IQR of 0.95--1.07. We applied the same HSC quality cuts to all 80 DESI galaxies with a fitted SDSS spectrum, each within $1^{\prime\prime}$ of a fitted SDSS or BOSS fiber (Table~\ref{tab:desi_population}). The 55 that pass have a median flux ratio of 0.973 and an IQR of 0.943--0.994. At the $3^{\prime\prime}$ diameter of SDSS spectrograph fibers, the same 49 galaxies without fitted SDSS spectra have a median flux ratio of 0.967 and an IQR of 0.939--1.004. Of the 55 galaxies with fitted SDSS spectra, 54 pass the mask-coverage cut at this aperture. Their median flux ratio is 0.967 with an IQR of 0.934--0.999. For the galaxies without fitted SDSS spectra, the change in median flux ratio between apertures shows that the near-unity agreement in the larger aperture can depend on the spatial scale being measured.

Figure~\ref{fig:hsc_operator} shows that the agreement extends beyond aperture flux to the broad spiral-arm patterns of both galaxies. In the bottom row, the arms are difficult to distinguish in the blurred SDSS image, yet the prediction reproduces the pattern seen independently by HSC. The same prediction contains a straight streak. This artifact arises from elongated Gaussian components in the fitted field and is more pronounced in the example shown in Figure~\ref{fig:limitations}(a). HSC's sharper PSF and finer pixel sampling also reveal knots and core structure that are not fully reproduced, on scales less well constrained by the lower-resolution SDSS observations used to fit the field. Section~\ref{sec:discussion} examines these artifacts and what the fitted observations constrain below their seeing scale.

\subsubsection{A new spectrograph: DESI}
\label{sec:desi}

We further test operator transfer by querying the field through a DESI spectrograph, which has a smaller fiber and a different spectroscopic response. Of the 479 DESI spectra in our sample, 250 (52~percent) combine multiple exposures. For these coadded spectra, we predict each exposure separately using its recorded fiber position and angular aperture before combining the predictions. The seeing is represented by a Moffat profile with $\beta=3.5$ \citep{guy2023desipipeline}. We use the seeing recorded for each exposure when available and otherwise adopt the $1.1^{\prime\prime}$ default of the \href{https://github.com/desihub/desispec/blob/0.51.13/py/desispec/fiberfluxcorr.py\#L16}{DESI point-source flux correction routine}. DESI's calibration accounts for point-source light falling outside the fiber \citep{guy2023desipipeline}, so we apply the corresponding correction to our predictions. The exposure predictions are then combined, the coadd resolution matrices applied on DESI's wavelength grid, and the fluxes converted to AB.

Targets are selected from DESI DR1 using its galaxy, stellar, and quasar classifications, with $0.005<z<0.6$ for galaxies. We rank eligible targets by observed SDSS $r$-band aperture flux, so the samples below are deliberately bright and test operator transfer rather than represent the full DESI populations. For the samples without fitted SDSS spectra, we select the brightest 80 targets in each class more than $10^{\prime\prime}$ from any fitted SDSS or BOSS fiber. Stars and quasars in these samples also have no fitted fiber within their observed image footprints. We exclude one quasar without a fitted SDSS spectrum whose redshift DESI flags as uncertain, leaving 79. For the samples with fitted SDSS spectra, we select the brightest 80 eligible targets in each class with a fitted SDSS or BOSS fiber within $1^{\prime\prime}$. Three stellar candidates failed the DESI redshift or wavelength-coverage cuts, so we used the next three eligible stars in brightness order.

Table~\ref{tab:desi_population} summarizes the flux ratio and shape error for our sample. The flux ratio is the summed predicted flux divided by the summed observed flux over valid DESI pixels from 4000 to 9000\,\AA, which lies within the wavelength coverage shared by the SDSS and DESI spectrographs. For the shape comparison, we average the observed and predicted spectra in $100$\,\AA\ bins using DESI-provided inverse-variance weights, giving greater weight to wavelengths with smaller observational uncertainties. The shape error measures the remaining root-mean-square (RMS) difference after removing the overall flux level, which the flux ratio already measures, as a percentage of the median absolute observed bin flux. Figure~\ref{fig:desi_best} in Appendix~\ref{app:desi_spectra} shows the six spectra in each sample without fitted SDSS spectra whose predictions agree most closely with DESI.

\begin{table}
    \centering
    \caption{Predictions of selected DESI DR1 spectra from the fixed SDSS-fitted field. $N$ is the number of spectra. Flux ratios compare predicted and observed flux summed over valid pixels from 4000 to 9000\,\AA. The median ratio and IQR are reported. The shape error is the RMS residual in $100$\,\AA\ bins with the overall flux level removed, divided by the median absolute observed bin flux. Its median over spectra is given in percent. In the Spectrum column, `Present' denotes an SDSS or BOSS spectrum included in the fit within $1^{\prime\prime}$ of the target while `Absent' denotes no fitted SDSS or BOSS fiber within $10^{\prime\prime}$.}
    \label{tab:desi_population}
    % Generated by scratch/paper_writing/5.2/make_desi_table.py; do not hand-edit.
% Source SHA256: 55b4e9fe1da279ab170daedda23d2c30c1b95e324c963ac003958889ce747388
\begingroup
\sisetup{detect-all,group-digits=false}
\setlength{\tabcolsep}{3.5pt}
\renewcommand{\arraystretch}{1.12}
\begin{tabular*}{\columnwidth}{@{\extracolsep{\fill}}llS[table-format=2.0]S[table-format=1.3]lS[table-format=2.1]@{}}
\toprule
Class & Spectrum & {$N$} & {Median ratio} & \multicolumn{1}{c}{IQR} & {Shape (\%)} \\
\midrule
\multirow{2}{*}{Galaxies} & Absent & 80 & 0.986 & $[0.924,\,1.019]$ & 9.9 \\
 & Present & 80 & 0.932 & $[0.850,\,1.017]$ & 5.6 \\
\addlinespace[2pt]
\midrule[0.2pt]
\addlinespace[2pt]
\multirow{2}{*}{Quasars} & Absent & 79 & 0.974 & $[0.852,\,1.213]$ & 22.8 \\
 & Present & 80 & 0.993 & $[0.790,\,1.191]$ & 14.7 \\
\addlinespace[2pt]
\midrule[0.2pt]
\addlinespace[2pt]
\multirow{2}{*}{Stars} & Absent & 80 & 0.814 & $[0.678,\,0.915]$ & 16.6 \\
 & Present & 80 & 0.988 & $[0.924,\,1.027]$ & 4.6 \\
\bottomrule
\end{tabular*}
\endgroup

\end{table}

The samples with fitted SDSS spectra have smaller median shape errors than the separately selected samples without them in all three classes (Table~\ref{tab:desi_population}). For galaxies, the median shape error is 5.6~percent for those with fitted SDSS spectra and 9.9~percent for those without them, while the median integrated-flux ratio is closer to unity for galaxies without fitted SDSS spectra (0.986 versus 0.932). Quasar flux ratios span a wide range in both samples, with median shape errors of 14.7~percent for quasars with fitted SDSS spectra and 22.8~percent for those without them. The stellar predictions for sources without fitted SDSS spectra are systematically faint, with a median flux ratio of 0.814 compared with 0.988 for stars with fitted SDSS spectra. The DESI observations were taken about two decades after the SDSS imaging \citep{desicollaboration2026dr1}, and we apply no variability or proper-motion correction. Section~\ref{sec:discussion} examines the brightness dependence of the flux deficit among stars without fitted SDSS spectra.

\begin{figure*}
    \centering
    \includegraphics[width=0.9\textwidth]{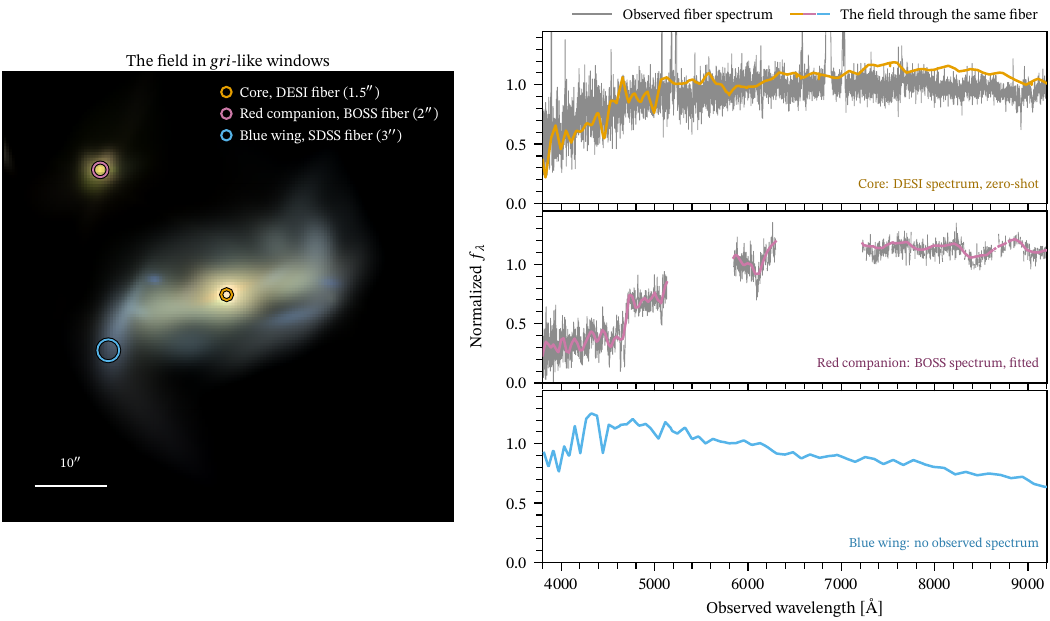}
    \caption{\textbf{Virtual fiber spectra around the spiral galaxy in the top row of Figure~\ref{fig:hsc_operator}.} The color image and all three spectral predictions are rendered from the field. Circles mark the apertures, with matching colors for the predicted spectra on the right. Gray curves show the observed spectra. The top panel compares the zero-shot prediction through a virtual DESI operator with the independent DESI spectrum. The middle panel shows the fitted BOSS spectrum of the background galaxy. The bottom panel predicts a $3^{\prime\prime}$-diameter SDSS fiber spectrum on the blue spiral arm, where no observed spectrum is available. Spectra are normalized for display. Because the field is continuous in position and wavelength, a virtual aperture can be placed anywhere in the fitted field to predict a spectrum, as illustrated by the blue wing.}
    \label{fig:virtual_ifu}
\end{figure*}

Figure~\ref{fig:virtual_ifu} compares three apertures in the region around the galaxy in the top row of Figure~\ref{fig:hsc_operator}. At the core, the prediction follows the broad DESI continuum with an integrated-flux ratio of 0.999 and a shape error of 12.8~percent, although it remains redder and does not reproduce the narrow emission peaks. The nearby source labeled ``Red companion'' is a background galaxy at $z=0.178$ with a fitted BOSS spectrum. The BOSS prediction agrees closely with the data, as the spectrum was fitted. On the blue wing, a virtual SDSS fiber of $3^{\prime\prime}$ diameter is rendered to showcase a spectrum where no observed spectrum is available. It is placed on the galaxy's extended spiral arm, about $18^{\prime\prime}$ from the core. Its broad color is constrained by the fitted images, while its finer spectral detail comes only from the shared dictionary.

\section{Discussion}
\label{sec:discussion}

Jointly fitting images and spectra to the same underlying model allows spectroscopic information to affect predictions beyond the fitted fibers. The comparison in Section~\ref{sec:withheld_spectra} demonstrates this for the 89 withheld spectra, where including spectroscopy substantially improves the predictions compared with images-only fits. The shared spectral dictionary allows spectra observed at a subset of positions to constrain the templates shared by every Gaussian component, regardless of whether a spectrum was observed at that point. This provides a way to connect sparse fiber measurements to the wider imaging coverage, creating an IFU-like field of predicted spectra across the entire imaged area. The blue-wing aperture in Figure~\ref{fig:virtual_ifu} demonstrates an application of this inherent multimodality by predicting a spectrum within an extended source where no observed spectrum is available. Its detailed shape therefore remains unvalidated, but the spectra fitted elsewhere in the field still constrain it through the shared dictionary.

Because the field is a spectral surface brightness on the SDSS flux scale, constructing virtual instrument operators allows the same fitted field to predict measurements through instruments that supplied no data to the fit. Section~\ref{sec:new_instruments} builds HSC and DESI operators and applies them to the field fitted to SDSS data alone. A common source model rendered through different PSFs is already used in forced photometry. The Farmer \citep{weaver2023farmer} fits each source with a model from The Tractor \citep{lang2016tractor} and can hold its shape and position fixed while fitting only its flux in each band. In our framework, the joint fit connects image pixels and fiber spectral bins, allowing both kinds of measurements to be predicted from the same field. Galaxies with no fitted SDSS spectrum within $10^{\prime\prime}$ have median flux ratios of 1.01 in $6^{\prime\prime}$-radius HSC apertures (49 galaxies; 0.967 at $3^{\prime\prime}$ diameter) and 0.986 for DESI spectra (80 galaxies), showcasing constraints that generalize beyond what the fitted data supply. Galaxies with a fitted SDSS or BOSS spectrum within $1^{\prime\prime}$ have a median DESI shape error of 5.6~percent, compared with 9.9~percent for galaxies without fitted SDSS spectra. Quasars and stars with fitted SDSS spectra also have lower median shape errors than those without them (Table~\ref{tab:desi_population}).

Joint reconstruction has been demonstrated for images with different PSFs and pixel scales \citep{joseph2021jsp,melchior2026scarlet2} and for multiband imaging combined with integral-field spectroscopy \citep{guilloteau2020fusion,marquis2026fusion}. In the JWST fusion of \citet{marquis2026fusion}, the spectral basis is derived from the observed NIRSpec cube. \citet{peng2026ifu} predict spatially resolved galaxy spectra with calibrated uncertainties by learning from a large sample of images and single-fiber DESI spectra. Our framework instead learns its spectral dictionary and spatial components jointly from the 400 exposures and 800 fiber spectra of a single sky region, requiring neither an IFU-derived spectral basis nor an external pretraining corpus.

Among the stars without fitted SDSS spectra in Table~\ref{tab:desi_population}, the median flux ratio is 0.814, and the shortfall grows with brightness. For the brightest third of the stars (Gaia $G\lesssim13.4$), total-flux estimates from a $6^{\prime\prime}$-radius aperture with a point-source correction give a median ratio to DESI of 0.73, compared with 1.01 in the faintest third. These bright stars also have broader profiles in the field than point sources, further reducing the fraction of their light collected by the DESI fiber. Gaia DR2 photometry provides a separate check on this brightness dependence \citep{gaia2016mission,gaia2018dr2}. We fit separate empirical color conversions for the field totals and DESI fluxes using the faintest third, then hold these conversions fixed for the brighter stars. From the faintest to the brightest third, the median DESI-to-Gaia ratio changes only from 1.005 to 1.039, whereas the field-to-Gaia ratio falls from 1.018 to 0.745, meaning the brightness-dependent deficit lies in the field's predicted total flux. Of the 27 stars in the brightest third, 25 have cores saturated in at least one fitted $g$, $r$, or $i$ exposure, and those pixels were masked from the fit (Section~\ref{sec:observations}). Brighter stars also tend to have more extensive saturation, so we cannot determine from this sample whether saturation is responsible for their flux deficit. Zero-shot fluxes for the brightest third of stars are therefore biased low in this field, because the fitted field underestimates their total light and spreads it more widely than a point source.

\begin{figure}
    \centering
    \includegraphics[width=0.95\columnwidth]{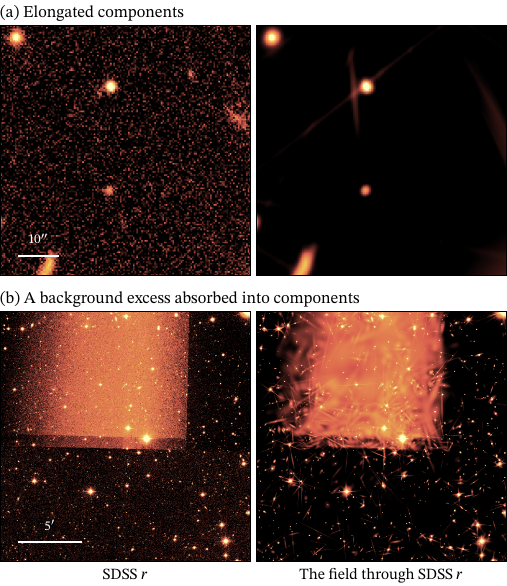}
    \caption{\textbf{Examples of known failure modes.} Artifacts in the fitted field. Each row compares the observed SDSS $r$-band image (left) with its prediction (right). In (a), elongated Gaussian components produce thin, straight streaks around the sources. Similar elongated artifacts are a known limitation of the original 3D Gaussian-splatting method that we adapt here \citep{kerbl20233dgaussiansplattingrealtime}. In (b), SDSS exposures flagged as non-photometric were retained in the fit. With the survey background held fixed, their rectangular background excess is absorbed into the field as extended, filamentary structure.}
    \label{fig:limitations}
\end{figure}

Our representation has spatial degrees of freedom below the SDSS seeing, but recovering that structure from the blurred observations is an ill-posed inverse problem. Of the fitted components, 84.2~percent, carrying 74.6~percent of the $r$-band light, have intrinsic major-axis FWHM below $1.4^{\prime\prime}$, close to the $1.408^{\prime\prime}$ median PSF FWHM of the fitted images. The independent HSC images provide a qualitative check for two selected galaxies (Figure~\ref{fig:hsc_operator}), showing that the field reproduces broad spiral-arm patterns while knots and core structure are not fully reproduced. Figure~\ref{fig:limitations}(a) shows thin, straight streaks in the SDSS rendering that are absent from the data, resembling the elongated artifacts reported for Gaussian splatting \citep{kerbl20233dgaussiansplattingrealtime}. We tested an axis-ratio cut to remove these components, but an unconditional cut could also delete components carrying well-fitted source light, so we leave this problem for future work. Different intrinsic component distributions can produce similar predictions after PSF blurring, so the data constrain the final light profile through the instrument operator more directly than the underlying components that make it up. In this sub-PSF regime, the Gaussian basis, initialization, and density-control rules influence aspects of the component structure that the data constrain only weakly. We apply no explicit spatial or spectral prior to preserve the representation's flexibility to represent diverse morphologies and spectra within a common model. Structure below the SDSS seeing therefore requires independent validation before it can be interpreted as resolved morphology \citep{lanusse2021deepgenerative}.

This intended flexibility allows errors in the survey inputs, such as the background estimates, to become structure in the learned field. The survey background $B_i$ is held fixed, so broad components can absorb excess light left by an underestimated background. Figure~\ref{fig:limitations}(b) shows a block of SDSS $r$-band data that sits above its surroundings and is reproduced as extended, filamentary structure in the field. The affected SDSS exposures carry non-photometric calibration flags but were retained in the fit, and the fit consequently treats the residual background excess as genuine flux and adds it to the field. Our reconstruction therefore relies on the survey calibration and background estimates being sufficiently accurate for the adopted noise model. Scaling to larger survey areas will require careful quality control and data cleaning to identify and correct or exclude contaminated measurements before they enter the fit.

The fitted field rendered through fiber spectrograph operators can reproduce continuum shapes and integrated fluxes while also missing narrow features. Absorption troughs remain too shallow in the fitted and withheld stellar examples, and narrow emission peaks are broadly smoothed over. The coarse template grid may contribute: its knots are evenly spaced in log wavelength, giving a separation of about 84\,\AA\ at 6000\,\AA, for example, compared to an instrumental LSF width of only a few angstroms (Section~\ref{sec:fiber_operator}). Other grid spacings and definitions have not been tested. The continuum and integrated-flux agreement in Sections~\ref{sec:fitted_field} and~\ref{sec:synthesis} therefore does not establish accurate line strengths or widths.

Our comparisons measure prediction errors for selected observed samples, but we do not yet provide calibrated uncertainties for an individual flux or spectrum queried from the field. Estimating those uncertainties requires accounting for the different combinations of spatial components and shared spectral templates consistent with the data. Component degeneracies also complicate error estimation in scarlet \citep{melchior2018scarlet}. Covariance approximations or posterior sampling, discussed for Jolideco by \citet{donath2024jolideco}, are possible approaches to providing uncertainties on measurements derived from the fitted field.

\section{Conclusions}
\label{sec:conclusions}

We have presented Spectro-Spatial Survey Fusion, a framework for jointly fitting a continuous spectral surface-brightness field to images and fiber spectra through their instrument responses. An adaptive Gaussian basis and a shared spectral dictionary are learned together from the observations, allowing the same model to describe galaxies, stars, and quasars without prescribing a separate profile for each source. We applied this representation to 400 SDSS exposures and 800 fiber spectra over an $84^{\prime}\times84^{\prime}$ region of COSMOS. The field reproduces the spatial structure and broad spectral continua of the sources, with 96~percent of the fitted spectra agreeing in integrated flux to within ten percent, although narrow spectral features are reproduced less accurately.

Our framework represents the underlying sky as a continuous spectral surface-brightness field, allowing the same fitted model to predict measurements through different instrument operators. We use this field to predict withheld fiber spectra, HSC images, and DESI spectra without refitting. Including spectroscopy lowers the mean squared pull relative to an images-only control for 88 of the 89 withheld spectra at each of three random seeds. In the selected galaxy samples with no fitted SDSS fiber within $10^{\prime\prime}$, the median predicted-to-observed flux ratios are 1.01 in $6^{\prime\prime}$-radius HSC $g$-band apertures (0.967 at $3^{\prime\prime}$ diameter) and 0.986 for integrated DESI spectra.

The predictions through new instrument responses motivate joint fits of complementary surveys. Including higher-resolution imaging such as HSC would add direct constraints on fine spatial structure, alongside the wavelength information supplied by spectroscopy. Extending the fit to larger sky regions would test the representation across a wider range of sources and observing conditions. This framework offers a route toward a continuous model of the sky built from the combined spatial and spectral information of many surveys, where the detail resolved by one instrument can help constrain what remains unresolved by another.

\section*{\texorpdfstring{Acknowledgements}{Acknowledgements}}

This research is supported by a grant from Alex Gerko, Founder of XTX Markets.

ChatGPT and Codex (OpenAI) and Claude (Anthropic) were used to assist with code development, analysis, and manuscript outlining and editing. The authors take full responsibility for the content and conclusions of this work.

\section*{\texorpdfstring{Data availability}{Data availability}}

The SDSS, HSC PDR3, DESI DR1, and Gaia DR2 data are available through their respective public archives. A frozen snapshot of the research code and supporting metadata is available at \url{https://github.com/GeorgeVassilakis/survey-fusion-paper-code}. This unmaintained release excludes fitted checkpoints and the complete HSC/DESI analysis workflows.

\bibliographystyle{mnras}
\bibliography{references}

\clearpage
\appendix
\raggedbottom
{
\section{Notation}
\label{app:notation}

Table~\ref{tab:notation} lists the symbols used in this paper and the section that defines each.

\noindent\begin{minipage}{\columnwidth}
    \makeatletter
    \def\@captype{table}
    \let\@makecaption\SFB@maketablecaption
    \makeatother
    \centering
    \caption{Summary of notation.}
    \label{tab:notation}
    \small

    \begin{tabular}{@{}lp{0.62\columnwidth}l@{}}
        \toprule
        Symbol & Meaning & Section \\
        \midrule
        $\mathcal{X}$, $\boldsymbol{x}$ & fitted region and position in flat-sky coordinates & \ref{sec:measurement_model} \\
        $\Lambda$, $\lambda$ & modeled wavelength range and observed wavelength & \ref{sec:measurement_model} \\
        $F_\lambda(\boldsymbol{x},\lambda)$ & spectral surface-brightness field & \ref{sec:measurement_model} \\
        $i$, $\mathcal{D}$ & datum index and set of all image pixels and spectral bins & \ref{sec:measurement_model} \\
        $\mathcal{D}_{\mathrm{img}}$, $\mathcal{D}_{\mathrm{spec}}$ & image pixels and spectral bins & \ref{sec:measurement_model} \\
        $e(i)$ & exposure or spectrum containing datum $i$ & \ref{sec:measurement_model} \\
        $d_i$, $\widehat{d}_i$ & background-subtracted measurement and prediction & \ref{sec:measurement_model} \\
        $B_i$, $\mathcal{A}_i$, $\epsilon_i$ & survey background, observation operator, and noise & \ref{sec:measurement_model} \\
        $j$, $J$ & component index and number of components & \ref{sec:components} \\
        $A_j$, $G_j$ & component amplitude and unit-normalized spatial Gaussian & \ref{sec:components} \\
        $\boldsymbol{\mu}_j$, $\boldsymbol{\Sigma}_j$ & center and covariance of $G_j$ & \ref{sec:components} \\
        $s_j$ & component spectrum & \ref{sec:components} \\
        $k$, $K$, $T_k$ & template index, count, and shape & \ref{sec:components} \\
        $w_{jk}$, $\zeta_j$ & template coefficients and log-wavelength shift & \ref{sec:components} \\
        $b(i)$, $R_b$ & band of image datum $i$ and system throughput of band $b$ & \ref{sec:image_operator} \\
        $\boldsymbol{x}_i$, $P^{\mathrm{pix}}_i$ & pixel-center position and pixel-convolved PSF & \ref{sec:image_operator} \\
        $\overline{s}_{bj}$ & band average of $s_j$ & \ref{sec:image_operator} \\
        $D_e$, $\theta^{\mathrm{fib}}_e$ & fiber aperture and radius & \ref{sec:fiber_operator} \\
        $P_e$, $\sigma^{\mathrm{see}}_e$ & spectroscopic seeing PSF and standard deviation & \ref{sec:fiber_operator} \\
        $Q_{ej}$ & fraction of component $j$'s light collected by fiber $e$ & \ref{sec:fiber_operator} \\
        $\Lambda_i$, $L_i$ & wavelength interval and LSF of bin $i$ & \ref{sec:fiber_operator} \\
        $\overline{s}_{ij}$ & LSF-convolved bin average of $s_j$ & \ref{sec:fiber_operator} \\
        $\eta_e$ & point-source fraction admitted by fiber $e$ & \ref{sec:fiber_operator} \\
        $\sigma_i$, $\chi_i$ & observational uncertainty and pull & \ref{sec:held_out_design} \\
        $\mathrm{MSP}(\mathcal{C})$ & mean squared pull over $\mathcal{C}\subseteq\mathcal{D}$ & \ref{sec:held_out_design} \\
        $\boldsymbol{\theta}$ & model parameters & \ref{sec:objective} \\
        $\mathcal{D}^{\mathrm{fit}}_{\mathrm{img}}$, $\mathcal{D}^{\mathrm{fit}}_{\mathrm{spec}}$ & fitted image pixels and spectral bins & \ref{sec:objective} \\
        $v_i^{(t)}$, $\gamma_{\mathrm{spec}}$ & image variance at update $t$ and spectral weight & \ref{sec:objective} \\
        $g_i$, $\sigma^{\mathrm{det}}_i$ & effective gain and detector noise & \ref{sec:objective} \\
        $N_{\mathrm{ref}}$ & fixed normalization of the objective & \ref{sec:objective} \\
        \bottomrule
    \end{tabular}
\end{minipage}

\section{Optimization settings}
\label{app:optimization_settings}

The shared-template and sparse component optimizers both use momentum coefficients $(0.9,0.999)$ and $\epsilon=10^{-8}$. The initial learning rates in Table~\ref{tab:optimization_rates} follow the gsplat defaults mapped onto our parameter groups, with the position rate reduced tenfold. The position rate decreases exponentially to one percent of its initial value over the first 30{,}000 updates and is then held fixed. The other rates remain constant. We clip the combined gradient norm of the shared template parameters at unity, leaving per-component gradients unclipped. The implementation uses PyTorch \citep{paszke2019pytorch} with custom Triton kernels \citep{tillet2019triton} for rendering and sparse parameter updates.

\par\medskip
\noindent\begin{minipage}{\columnwidth}
    \makeatletter
    \def\@captype{table}
    \let\@makecaption\SFB@maketablecaption
    \makeatother
    \small

    \centering
    \caption{Initial learning rates for the canonical fit. Rates act on the stored parameters before exponentiation or softplus. Spatial coordinates and the template log-wavelength interval are each normalized to $[-1,1]$. Only the position rate decays.}
    \label{tab:optimization_rates}
    \begin{tabular}{lr}
        \hline
        Parameter & Initial learning rate \\
        \hline
        Position & $1.6\times10^{-5}$ \\
        Log width & $5\times10^{-3}$ \\
        Position angle & $10^{-3}$ \\
        Amplitude & $5\times10^{-2}$ \\
        Spectral mixture coefficients & $2.5\times10^{-3}$ \\
        Template values & $2.5\times10^{-3}$ \\
        Spectral shift & $1.25\times10^{-4}$ \\
        \hline
    \end{tabular}
\end{minipage}

\subsection{Component growth and pruning}
\label{app:component_lifecycle}

Growth is considered every 100 updates from update 600 through 14{,}900. At the first refinement, we set the growth threshold to the 70th percentile of the mean positional-gradient magnitudes among visited components and hold it fixed thereafter. At subsequent refinements, each mean is measured over the component's sampled visits since the previous refinement. Components above the threshold that are not already marked for removal are duplicated if their largest intrinsic standard deviation is at most three times the median PSF FWHM of the fitted images. Duplication retains the parent's position and widths. Larger candidates are replaced by two children with positions drawn from the parent Gaussian and both principal-axis widths divided by 1.6. Both operations copy the parent's spectral mixture coefficients and shift, with the amplitude division described in Section~\ref{sec:optimization}. There is no opacity-reset operation.

Pruning is evaluated on the same 100-update cadence and continues after growth ends. A component is removed if its estimated PSF-convolved peak is below $10^{-8}$ of the representative image noise in every band, or if its largest intrinsic standard deviation exceeds about $252^{\prime\prime}$. The peak estimate uses a circular Gaussian PSF with the median fitted-image FWHM in each band and includes the pixel-response variance. The noise reference is the median data-derived uncertainty of sampled unflagged pixels in those exposures. The prediction operator itself is unchanged by this approximation.

For the canonical fit, one image epoch comprises $\lfloor872{,}911{,}804/10^4\rfloor=87{,}291$ updates, giving 960{,}201 updates over 11 epochs. It reaches 5{,}639{,}839 components at update 14{,}900 and ends with 2{,}855{,}026 components. The last pruning event, at update 960{,}200, still removed 18 components.

\section{Zero-shot DESI spectra}
\label{app:desi_spectra}

Figure~\ref{fig:desi_best} shows the six DESI DR1 spectra in each sample without fitted SDSS spectra of Table~\ref{tab:desi_population} whose zero-shot predictions agree most closely with the observations. We rank each sample by the RMS difference of the $100$\,\AA\ bins of Section~\ref{sec:desi} without removing the overall flux level, so that a spectrum must match in both flux and shape. The panels show the closest agreement in each sample, and Table~\ref{tab:desi_population} gives the population medians.

\begin{figure*}
    \centering
    \includegraphics[width=\textwidth]{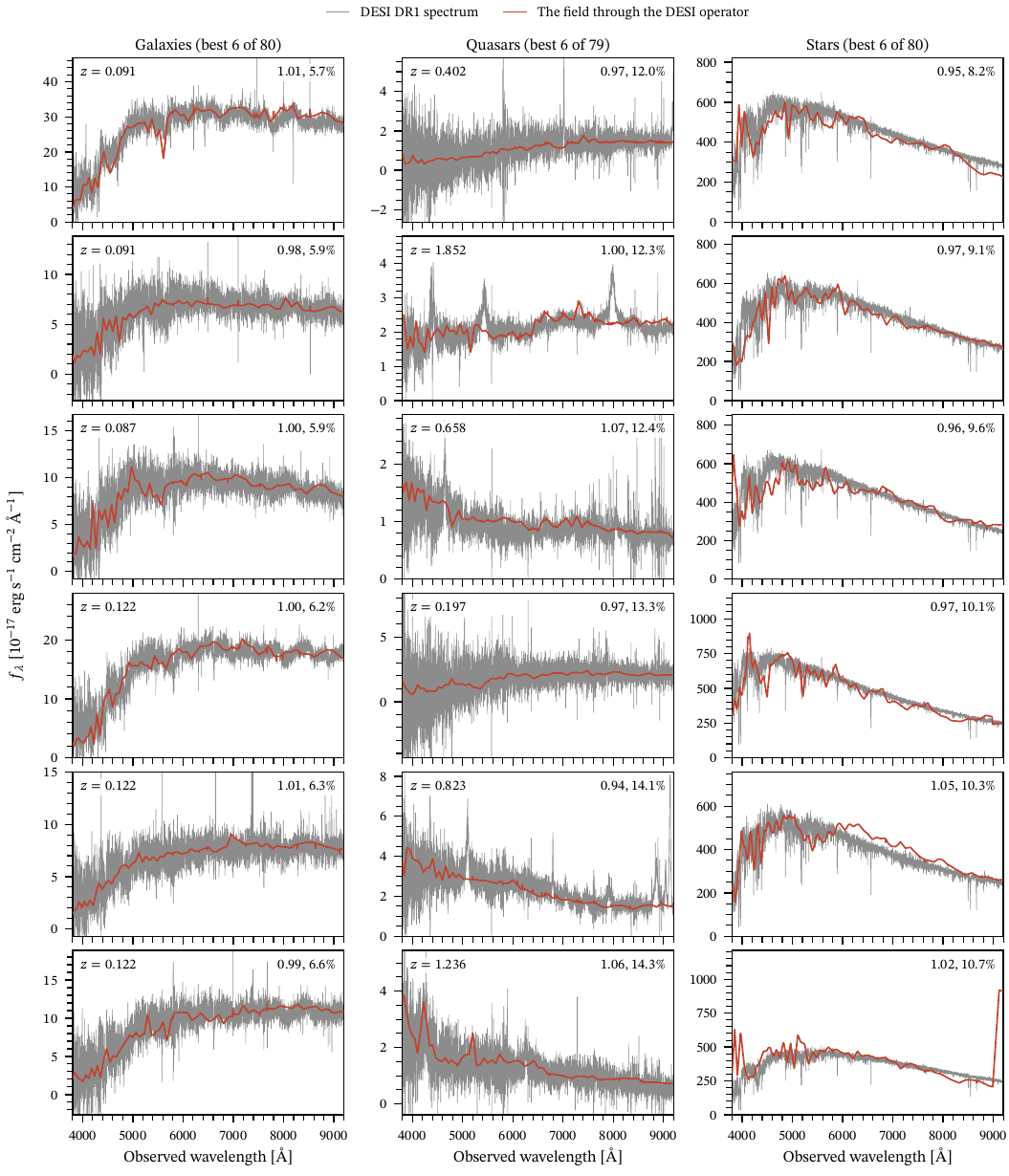}
    \caption{\textbf{Zero-shot DESI DR1 spectra for the six closest matches in each sample without fitted SDSS spectra.} No DESI data were used to fit the field or its shared spectral dictionary. The targets have only SDSS imaging constraints locally, while SDSS spectra fitted elsewhere inform their predictions through the shared dictionary. The columns show galaxies, quasars, and stars from left to right. Gray curves show the DESI observations, and red curves show the field rendered through the DESI operator at its fitted flux scale. Each panel gives the predicted-to-observed flux ratio and the RMS difference in $100$\,\AA\ bins over 4000--9000\,\AA, expressed as a percentage of the median absolute observed bin flux. Table~\ref{tab:desi_population} gives the medians for the corresponding samples without fitted SDSS spectra.}
    \label{fig:desi_best}
    \label{lastpage}
\end{figure*}
}

\bsp
\end{document}